\documentclass[reprint]{revtex4-1}
\usepackage{preamble}
\newcommand{\Yb}{$\text{YbMgGaO}_4$}

\begin{document}
\title{Finite-temperature behavior of a classical spin-orbit-coupled model for \Yb\ with and without bond disorder}
\date{\today}
\author{Edward Parker}
\email{tparker@alumni.physics.ucsb.edu}
\affiliation{Department of Physics, University of California, Santa Barbara, CA 93106}
\author{Leon Balents}
\affiliation{Kavli Institute for Theoretical Physics, University of California, Santa Barbara, CA 93106}

\begin{abstract}
We present the results of finite-temperature classical Monte Carlo simulations of a strongly spin-orbit-coupled nearest-neighbor triangular-lattice model for the candidate $\mathrm{U}(1)$ quantum spin liquid \Yb\ at large system sizes. We find a single continuous finite-temperature stripe-ordering transition with slowly diverging heat capacity that completely breaks the sixfold ground-state degeneracy, despite the absence of a known conformal field theory describing such a transition. We also simulate the effect of random-bond disorder in the model, and find that even weak bond disorder destroys the transition by fragmenting the system into very large domains -- possibly explaining the lack of observed ordering in the real material. The Imry-Ma argument only partially explains this fragility to disorder, and we extend the argument with a physical explanation for the preservation of our system's time-reversal symmetry even under a disorder model that preserves the same symmetry.
\end{abstract}

\maketitle

\section{Introduction}

The notion of the gapless $\text{U}(1)$ quantum spin liquid has come under extensive theoretical study in recent years \cite{Hermele04a, Hermele04b, LeeLee}. Such phases are similar in some ways to the more well-understood gapped quantum spin liquids \cite{Balents, Savary17} -- for example, they have topological-defect-like excitations with nontrivial braiding statistics -- but unlike in the gapped case, these excitations can be extended rather than localized, leading to quasi-long-range correlations similar to those found near a second-order phase transition or in quantum electrodynamics \cite{Savary12}. This paradigm may also be useful for understanding cuprate high-temperature superconductors \cite{Senthil}.

Unfortunately, conclusive evidence of experimental realizations of such a phase remain elusive. In many candidate $\text{U}(1)$ spin liquid materials, the $\text{U}(1)$ gauge field appears to be coupled to fractionalized spinon excitations that form a Fermi surface at low temperature \cite{Motrunich05, Motrunich06, Galitski, Lee}. The strongly spin-orbit-coupled material \Yb\ has recently come under both theoretical \cite{Y-DLi16a, Y-DLi17a, Y-DLi17b} and experimental \cite{YLi15a, YLi15b, YLi16, Shen, Paddison, Toth, YLi17b} study as a potential $\text{U}(1)$ spin liquid with such a spinon Fermi surface. The material shows low-temperature magnetic heat capacity $C_v \approx T^{2/3}$ \cite{YLi15b} in good agreement with theoretical predictions \cite{Motrunich05}, no sign of magnetic ordering down to $48$ mK (despite a spin exchange interaction energy of about $4$ K) \cite{YLi16}, and a continuum of magnetic excitations \cite{Shen, Paddison}, which support this picture.

However, recent density-matrix renormalization group (DMRG) simulations suggest that a perfectly clean sample of \Yb\ should show stripe order at zero temperature \cite{Luo, Zhu17}. Ref.~\onlinecite{Zhu17} found that adding explicit disorder to the model destroys the ground-state ordering, and proposed that the experimentally observed lack of ordering and other spin-liquid-like effects actually arise from sample disorder.

In this article, we use classical Monte Carlo simulations to extend this line of argument to the finite-temperature regime. We first perform finite-temperature Monte Carlo simulations of the classical version of a widely used microscopic model for \Yb\ in order to establish the existence, nature, and experimental signatures of the finite-temperature ordering transition(s). We then consider the effect of adding explicit disorder to the Hamiltonian via a model in which the disorder strength can be tuned continuously. Similarly to Ref.~\onlinecite{Zhu17}, we find that even very weak disorder removes any finite-temperature phase transition -- even if the disorder model preserves time-reversal symmetry, which could logically still be broken spontaneously. We propose that, at least at the classical level, even infinitesimal disorder removes the ground state's long-range order. This result supports our proposed disorder model as being realistic for \Yb, and provides further evidence that effects less exotic than spin-liquid physics may suffice to explain the lack of experimentally measured ordering in the material.

The article is organized as follows. We introduce the microscopic model below. Sec.~\ref{Classical GS} (largely summarizing previous work) discusses the model's classical zero-temperature phase diagram and shows that there are six degenerate ground states related by $\mathbb{Z}_6$ symmetry. Sec.~\ref{OP} discusses our choice of order parameter for measuring the sixfold symmetry breaking. Sec.~\ref{Clean MC} presents the results of our classical Monte Carlo simulations on the clean model. We find only a single, continuous symmetry-breaking transition at moderate temperature. We present a theoretical analysis of this transition, but are unable to determine its universality class. Sec.~\ref{Disordered MC} presents the results of our simulations when we incorporate explicit disorder into the model. We find that infinitesimal disorder destroys the phase transition, and offer an explanation based on a generalization of the Imry-Ma argument. Sec.~\ref{Conclusion} concludes.

\subsection{The model}

\begin{figure}
\includegraphics[page=1]{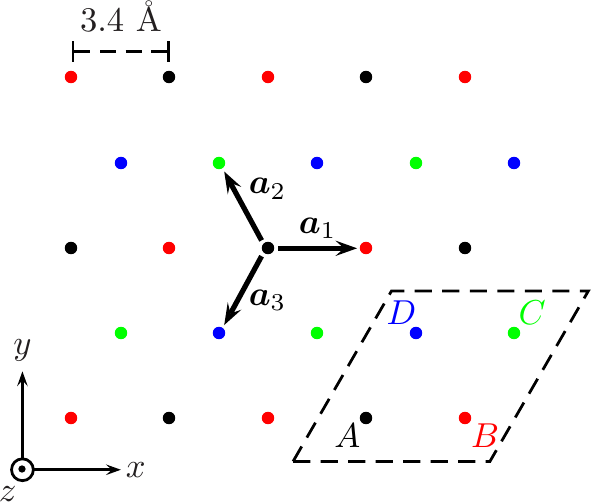}
\caption{\label{Lattice} The magnetic $\text{Yb}^{3+}$ ions form a triangular lattice with a lattice spacing of $3.4$ \AA. The nearest-neighbor bonds are oriented parallel to the three directions $\bm{a}_i$. The four sublattices $A$, $B$, $C$, $D$ described in the main text are colored differently.}
\end{figure}

\Yb\ consists of structurally clean layers of magnetic $\text{Yb}^{3+}$ ions and nonmagnetic oxygen atoms separated at $8.4$ \AA\ by disordered $\text{Mg}^{2+}$, $\text{Ga}^{3+}$, and $\text{O}^{2-}$ ions. The $\text{Yb}^{3+}$ ions in each layer form a structurally clean triangular lattice (illustrated in Fig.~\ref{Lattice}) with a lattice spacing of $3.4$ \AA, so we can neglect interlayer interactions and treat the system as two-dimensional \cite{YLi15a}.  We set the lattice spacing equal to $1$ throughout. The $4f$ electrons in the valence shell are quite localized, so nearest-neighbor interactions should dominate. (Recent experiments suggest that second-neighbor interactions may be nonnegligible \cite{Paddison, Zhang}, but we do not consider such interactions in this work.) As with most rare-earth magnets, spin-orbit (SO) effects are very important and the exchange couplings are strongly anisotropic in both spin and real space \cite{Y-DLi16b}.  Each $\text{Yb}^{3+}$ ion has $13$ electrons in the $4f$ valence shell, but the strong spin-orbit interaction breaks the $14$-fold energy degeneracy given by Hund's rule down to an eightfold degeneracy, and the electric crystal field (with point group $D_\text{3d}$) further splits the energy levels down to twofold-degenerate Kramer's doublets, which are separated by a large energy gap of $420$ K.  The electronic degrees of freedom can therefore be well approximated by a pseudospin-$1/2$ (as confirmed by low-temperature magnetic entropy measurements of $\ln 2$ per spin) \cite{YLi15b}.

The system has the space symmetry group $\text{R} \bar{3} \text{m}$, with generators $T_1$, $T_2$, $C_3$, $C_2$, and $\mathcal{I}$ described in Ref.~\onlinecite{Y-DLi16a}. We will only consider the system in the absence of external magnetic fields, and assume time-reversal symmetry as well.  Ref.~\onlinecite{YLi15a} found that the most generic nearest-neighbor spin Hamiltonian invariant under this symmetry group is
\begin{align}
H &= \sum_{\langle i j \rangle} \Big[ J_{zz} S_i^z S_j^z + J_\pm \left( S_i^+ S_j^- + S_i^- S_j^+ \right) \label{Ham} \\
&\hspace{33pt} + J_{\pm \pm} \left( \gamma_{ij} S_i^+ S_j^+ + \gamma_{ij}^* S_i^- S_j^- \right) \nn \\
& - \frac{1}{2} i J_{z \pm} \Big( \left( \gamma_{ij}^* S_i^+ - \gamma_{ij} S_i^- \right) S_j^z + S_i^z \left( \gamma_{ij}^* S_j^+ - \gamma_{ij} S_j^- \right) \Big) \Big], \nn
\end{align}
where $\langle i j \rangle$ denotes nearest-neighbor sites and $S^\pm := S^x \pm i S^y$ \cite{Zhu17supp}. The phase factor $\gamma_{ij} = \gamma_{ji}$ equals $1$, $e^{i 2 \pi / 3}$, and $e^{-i 2 \pi / 3}$ for bonds parallel to the directions
\begin{align*}
\bm{a}_1 = \left( 1, 0, 0 \right), \hspace{5pt} \bm{a}_2 = \left( -\frac{1}{2}, \frac{\sqrt{3}}{2}, 0 \right), \hspace{5pt} \bm{a}_3 = \left( -\frac{1}{2}, -\frac{\sqrt{3}}{2}, 0 \right)
\end{align*}
illustrated in Fig.~\ref{Lattice}, respectively. The spatially isotropic $J_{zz}$ and $J_\pm$ terms constitute the usual $XXZ$ model, while the $J_{\pm\pm}$ and $J_{z\pm}$ terms constitute the spin-orbit (SO) interaction (for which Refs.~\onlinecite{Zhu17supp, Zhu18supp} give a geometric interpretation).

Thermodynamic measurements suggest that for \Yb, $J_{zz} = 0.99$ K and $J_\pm = 0.91$ K. The SO couplings are more difficult to estimate experimentally, but electron spin resonance measurements suggest that $J_{\pm\pm} = \pm 0.15$ K and $J_{z \pm} = \pm 0.05$ K (the measurements cannot determine the signs of the couplings) \cite{YLi15a}. The Hamiltonian \eqref{Ham} should also be a realistic model for several other triangular antiferromagnets with strong spin-orbit coupling, such as the rare-earth materials $\text{RCd}_3 \text{P}_3$, $\text{RZn}_3 \text{P}_3$, $\text{RCd}_3 \text{As}_3$, $\text{RZn}_3 \text{As}_3$, and $\text{RO}_2 \text{CO}_3$ \cite{Y-DLi16a}.

\section{Classical ground states \label{Classical GS}}
\begin{figure}
\includegraphics[width=\columnwidth]{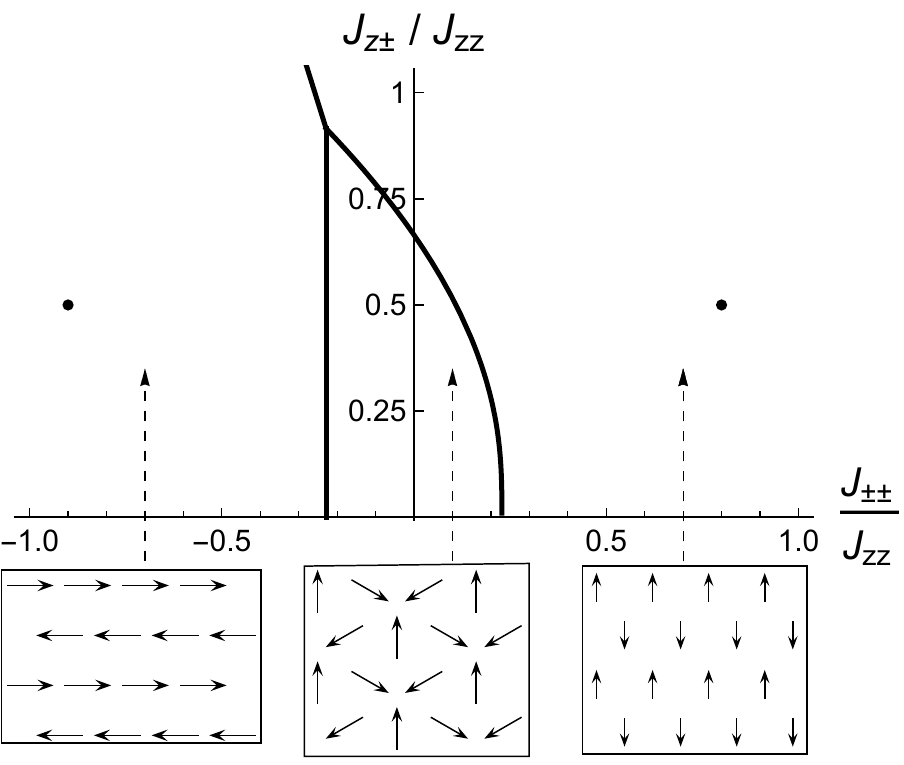}
\caption{Cross-section with $J_\pm = 0.915 J_{zz}$ of the approximate $T = 0$ phase diagram for the Hamiltonian \eqref{Ham} from the Luttinger-Tisza ansatz. In the strong SO regime $|J_{\pm\pm}| \gg 0$, there are six degenerate stripe-ordered ground states. In the ``in-plane SO phase'' with $J_{\pm\pm} \ll 0$, the spins lie in the $x$-$y$ plane and point along the stripes. In the ``out-of-plane SO phase'' with $J_{\pm\pm} \gg 0$ (and extending to weak negative $J_{\pm\pm}$ for large $J_{z \pm}$), they point perpendicular to the stripes and tilt out of the $x$-$y$ plane (and so appear shorter when viewed from above). For weak SO couplings $J_{\pm\pm}, J_{z\pm} \ll J_{zz}$, the ground state shows in-plane $120^\degree$ order with an emergent continuous $U(1)$ symmetry in spin space. The two black dots in the SO phases represent the Hamiltonians whose finite-temperature behavior is investigated in this paper. \label{GS phase diagram}}
\end{figure}

Finding even the classical ground state of the Hamiltonian \eqref{Ham} is highly nontrivial. Switching the sign of $J_{z \pm}$ is equivalent to a $180^\degree$ rotation about the $z$-axis in spin space, so we need only consider $J_{z \pm} \geq 0$ \cite{Y-DLi16a}. A powerful tool for finding the exact ground states of frustrated classical spin systems is the Luttinger-Tisza theorem \cite{Luttinger}. Unfortunately, the theorem is only guaranteed to work when the spins lie on a Bravais lattice and there is at least a continuous $\text{U}(1)$ spin symmetry \cite{Liu}, and the latter condition fails for our system. Nevertheless, the Luttinger-Tisza theorem does give the exact ground state over much of the phase diagram (see App.~\ref{GS app} for more details), allowing us to create an approximate phase diagram illustrated in Fig.~\ref{GS phase diagram} with three phases:
\begin{itemize}
\item If the SO couplings $J_{\pm\pm}$ and $J_{z\pm}$ are zero, then the system reduces to the $XXZ$ model, which has a $\text{U}(1)$ spin symmetry. The estimates for $J_{zz}$ and $J_{\pm}$ given above indicate that \Yb\ is in the easy-plane regime, so the Luttinger-Tisza theorem gives that the ground states form a continuous set of in-plane $120^\degree$-order states. Slightly away from $XXZ$ point, the Luttinger-Tisza theorem fails. But the energy of the $120^\degree$ order state does not actually depend on the SO couplings, indicating that the $120^\degree$-order states remain exact ground states over some finite neighborhood of the $XXZ$ point. This phase therefore displays an emergent $\text{U}(1)$ symmetry, despite the Hamiltonian \eqref{Ham}'s only having discrete symmetry.
\item If $J_{\pm\pm}$ is sufficiently negative, then the spins lie in the $x$-$y$ plane and form stripes that run along the principal directions of the triangular lattice. We will refer to this phase as the ``in-plane SO phase.''
\item If $J_{\pm\pm}$ is not strongly negative, then the spins again form stripes. But in this case, they point perpendicular to the stripes and partially out of the plane. If the stripes run parallel to the $x$-axis, as illustrated in Fig.~\ref{GS phase diagram}, then the spins on one set of stripes lie in the $y$-$z$ plane in spin space, forming an acute angle with the $+y$-axis and an obtuse angle $\theta$ with the $+z$-axis, where
\beq
\tan(2 \theta) = \frac{J_{z\pm}}{J_{\pm \pm} + \frac{1}{4}(2 J_\pm - J_{zz})}, \quad \theta \in \left( \frac{1}{2} \pi, \pi \right). \label{theta}
\eeq
The spins on the other set of stripes point in the opposite direction. We will refer to this phase as the ``out-of-plane SO phase.''
\end{itemize}

The phase boundaries in Fig.~\ref{GS phase diagram} are not exact, due to the failure of the Luttinger-Tisza theorem in the regime of weak but nonzero SO couplings.  Ref.~\onlinecite{Y-DLi16a} performed Monte Carlo calculations on small ($6 \times 6$ and $12 \times 12$) systems and found good agreement. However, Ref.~\onlinecite{Liu} uses a more sophisticated generalization of the standard Luttinger-Tisza approach and Monte Carlo simulations of larger systems to argue for the existence of three small incommensurate phases (which are difficult to identify from small systems) near the phase boundaries illustrated in Fig.~\ref{GS phase diagram}, each characterized by multiple incomensurate ordering wavevectors. In this work, we will not be concerned with the exact details of the classical zero-temperature phase diagram (which is of course of dubious experimental relevance for \Yb, which probably has strong quantum fluctuations). Instead, we will consider finite-temperature phase transitions, where quantum fluctuations should be negligible, for couplings deep in the two stripe-ordered spin-orbit phases (indicated by the two black dots in Fig.~\ref{GS phase diagram}), where the Luttinger-Tisza ansatz provably gives the exact classical ground state.

\section{SO phase order parameters \label{OP}}

In both SO phases, the six stripe-ordered ground states can be indexed by two integers: $b \in \{ 1, 2, 3\}$ specifying which principal lattice direction $\bm{a}_b$ the stripes run along, and $p \in \{ 0, 1\}$ specifying the spin orientations within each stripe. Explicitly, the spin at site $\bm{r}$ is
\beq
\bm{S}^{(b, p)}_{\bm{r}} = (-1)^p e^{i \bm{q}_b \cdot \bm{r}} \bm{n}_b = (-1)^{p + \Delta_b(\bm{r})} \bm{n}_b, \label{GS}
\eeq
where $\bm{q}_b := (2 \pi / \sqrt{3}) \hat{\bm{z}} \cross \bm{a}_b$ is the $M$ point at the center of the Brillouin zone edge parallel to $\bm{a}_b$ (App.~\ref{GS app}), and $\Delta_b(\bm{r}) := \left( 2 / \sqrt{3} \right) (\bm{a}_b \cross \bm{r})^z$ is an integer that numbers the stripe running in the $\bm{a}_b$-direction that contains the site $\bm{r}$. Time reversal changes the value of $p$ but not $b$. In the in-plane phase
\beq
\bm{n}_b := \bm{a}_b, \label{IPn}
\eeq
and in the out-of-plane phase
\beq
\bm{n}_b := \cos(\theta) \hat{\bm{z}} \times \bm{a}_b + \sin(\theta) \hat{\bm{z}}, \label{OPn}
\eeq
where $\theta$ is given by \eqref{theta}.

\begin{figure}
\includegraphics[page=2]{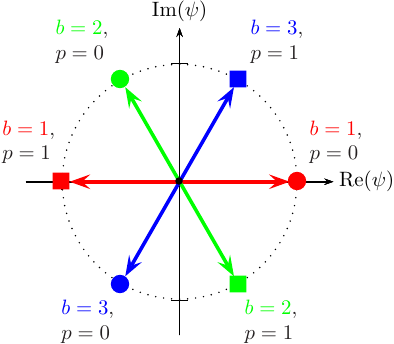}
\caption{The complex order parameter $\Psi$ defined in \eqref{Psi} equals a different sixth root of unity for each of the six stripe-ordered ground states in the SO phases. Values of $\Psi$ depicted in the same color correspond to ground state pairs related by time-reversal symmetry. \label{GSs}}
\end{figure}

This triangular spin-orbit-locked stripe order is fairly unusual, and the appropriate order parameter is not immediately obvious. Over each of the four triangular sublattices $A$, $B$, $C$, $D$ illustrated in Fig.~\ref{Lattice}, the spin orientations are uniform within each ground state and different between the six ground states.  So one possible order parameter is the average magnetization on a single sublattice, but we would like to incorporate the configuration of the spins on all four sublattices. To this end, consider the average dot product between two spin configurations
\[
\{ \bm{S} \} \cdot \{ \bm{S}' \} := \frac{1}{N} \sum_{\bm{r}} \bm{S}_{\bm{r}} \cdot \bm{S}'_{\bm{r}},
\]
where $N$ is the total number of spins.  The average dot product between an arbitrary spin configuration $\bm{S}$ and a ground state $\bm{S}^{(b, p)}$ given by \eqref{GS} is
\begin{align}
 \{ \bm{S} \} \cdot \big\{ \bm{S}^{(b, p)} \big\} &= (-1)^p \bm{n}_b \cdot \frac{1}{N} \sum_{\bm{r}} e^{i \bm{q}_b \cdot \bm{r}} \bm{S}_{\bm{r}} \nn \\
 &= (-1)^p \bm{n}_b \cdot \tilde{\bm{S}}(\bm{q}_b). \label{SdotSb}
\end{align}
Define $\omega := e^{i 2 \pi / 3}$, and consider the complex quantity
\beq
\Psi[\{ \bm{S} \}] := \sum_{b = 1}^3 \omega^{b-1} \{ \bm{S} \} \cdot \{ \bm{S}^{(b, 0)} \} = \sum_{b = 1}^3 \omega^{b-1} \bm{n}_b \cdot \tilde{\bm{S}}(\bm{q}_b). \label{Psi}
\eeq
For the six ground states, $\Psi$ takes on values over the sixth roots of unity (Fig.~\ref{GSs}):
\[
\Psi \big[ \big\{ \bm{S}^{(b, q)} \big\} \big] = (-1)^p \omega^{b - 1}.
\]
The symmetry transformations that relate the physical ground states are generated by $\mathcal{S}_6 := C_3^{-1} \mathcal{I}$, which rotates the system by $\pi/3$ about the $z$-axis and reflects across the $x$-$y$ plane.  The order parameter transforms as $\Psi \to e^{i \pi / 3} \Psi$ under this operation, so the ground state manifold has a $\mathbb{Z}_6$ symmetry that is clear from the figure.

We can express $\Psi$ as a spatial average over a local order parameter $\psi(x)$ by rewriting \eqref{Psi} as 
\beq
\Psi[\{ \bm{S} \}] = \frac{1}{N} \sum_{\bm{r}} \left[ \bm{S}_{\bm{r}} \cdot \left( \sum_{b = 1}^3 (-1)^{\Delta_b(\bm{r})} \omega^{b - 1} \bm{n}_b \right) \right]. \label{Psi2}
\eeq
If we formally consider the sum in large parentheses as a complex-valued spin configuration $\{ \bm{C} \}$, then this becomes $\Psi[\{ \bm{S} \}] = \{ \bm{S} \} \cdot \{ \bm{C} \}$. A straightforward calculation shows that $\bm{C}_{\bm{r}}$ is constant on each of the four sublattices, so $\Psi$ is a complex linear combination of the four sublattice magnetizations, with coefficients given in App.~\ref{OP app}. The on-site quantity in the large brackets above is not uniform within each ground state, so it is convenient to coarse-grain the spins to the $N/4$ plaquettes illustrated in Fig.~\ref{Lattice}, which contain one spin from each of the four sublattices. Denoting the location of a plaquette by $x$, we define
\beq
\psi(x) := \frac{1}{4} \sum_{s=A}^D \bm{C}_s \cdot \bm{S}_{x, s} \label{psi}
\eeq
so that $\Psi = 1/(N/4) \sum_x \psi(x)$. In each ground state $\psi(x) \equiv \Psi$ is uniform and equal to a sixth root of unity.

\section{Finite-temperature behavior of clean systems \label{Clean MC}}

Ref.~\onlinecite{Y-DLi16a} studied the finite-temperature behavior of the classical model \eqref{Ham} by performing classical Monte Carlo simulations on very small ($6 \times 6$ and $12 \times 12$) systems over a wide region of the phase diagram.  But because the spins in the classical model \eqref{Ham} are $\text{O}(3)$ vectors whose orientations can vary continuously, the model is very sensitive to thermal fluctions and requires very large systems and many Monte Carlo sweeps to equilibrate to the thermodynamic limit \cite{PricePerkins13}. In this section, we report results of classical Monte Carlo simulations for only two sets of couplings -- one deep in each of the SO phases of the $T = 0$ phase diagram -- but on much larger systems than those studied in Ref.~\onlinecite{Y-DLi16a}, and with much finer temperature resolution.

\subsection{Monte Carlo simulation results}

We chose the couplings
\[
J_{zz} = 1,\ J_\pm = 0.915,\ J_{\pm\pm} = -0.9,\ J_{z\pm} = 0.5
\]
and
\[
J_{zz} = 1,\ J_\pm = 0.915,\ J_{\pm\pm} = +0.8,\ J_{z\pm} = 0.5
\]
as representative points in the in-plane and out-of-plane SO phases, respectively. ($J_{zz} > 0, J_\pm = 0.915 J_{zz}$ represent experimentally realistic values for \Yb\, based on thermodynamic measurements, but we chose SO coupling values significantly larger than those suggested by experiment in order to isolate the physical effects of strong SO interactions, as they are less well-understood than the $XXZ$ couplings'. Furthermore, DMRG calculations suggest that quantum fluctuations may stabilize the stripe order in \Yb, despite its SO couplings' being relatively weak from a classical standpoint \cite{Zhu17}.) The critical behavior of the systems in both $T = 0$ phases turns out to be fairly similar. For each set of couplings, we performed a detailed sweep of the temperature range $1.0 < T < 2.0$ on $L \times L$ systems with periodic boundary conditions for $L = 46$, $64$, $90$, and $128$. We used the Metropolis Monte Carlo algorithm with $4\mbox{--}10 \times 10^5$ equilibration sweeps followed by $1\mbox{--}5 \times 10^6$ measurement sweeps, performing more sweeps near the critical temperature to compensate for critical slowing down.

\subsubsection{Bulk thermodynamic quantities \label{Thermo}}

\begin{figure*}
\includegraphics[width=0.47\linewidth]{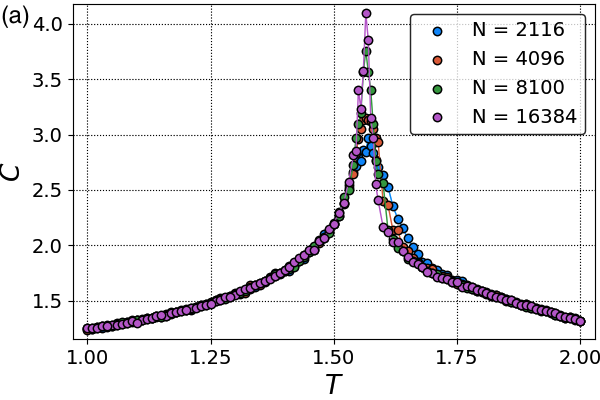} \hspace{10pt} \includegraphics[width=0.47\linewidth]{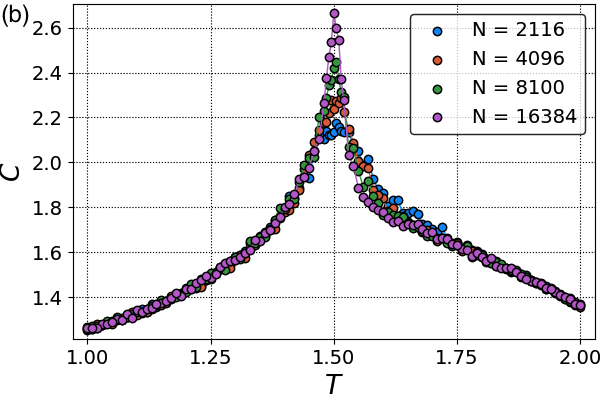}
\caption{Heat capacity per site for clean systems of various sizes in the (a) in-plane and (b) out-of-plane phase. \label{CleanCs}}
\end{figure*}

\begin{figure*}
\includegraphics[width=0.47\linewidth]{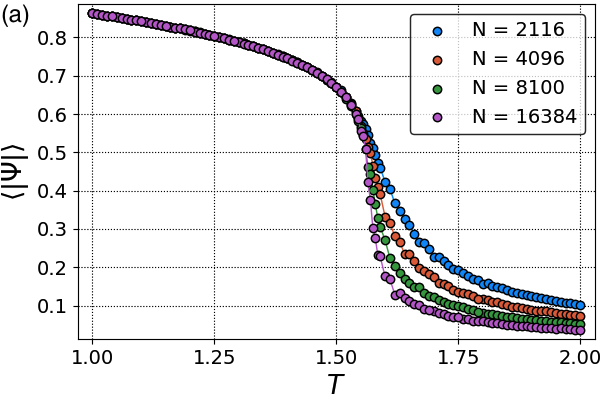} \hspace{10pt} \includegraphics[width=0.47\linewidth]{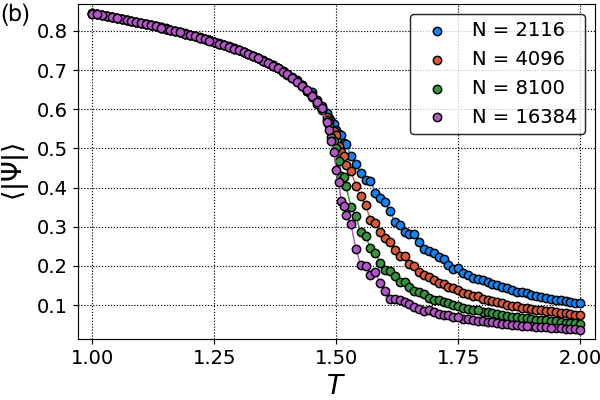}
\caption{Order parameter \eqref{Psi} for clean systems of various sizes in the (a) in-plane and (b) out-of-plane phase. \label{CleanOPs}}
\end{figure*}

\begin{figure*}
\includegraphics[width=0.19\linewidth]{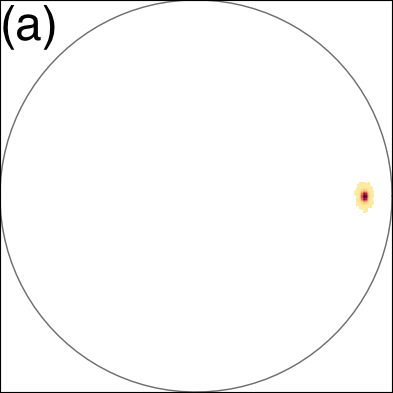} \includegraphics[width=0.19\linewidth]{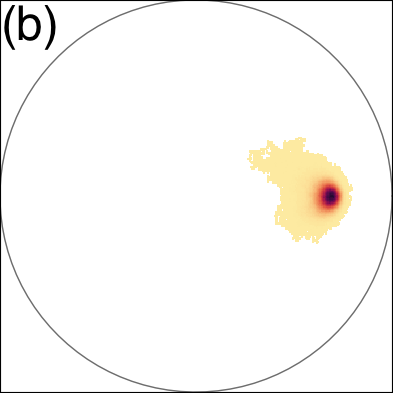} \includegraphics[width=0.19\linewidth]{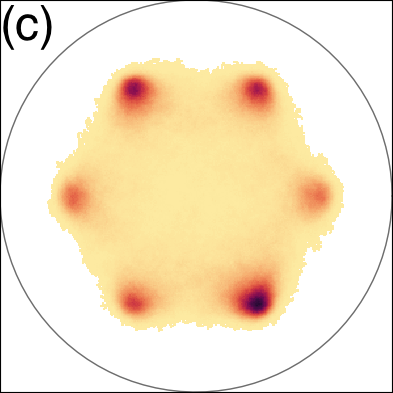} \includegraphics[width=0.19\linewidth]{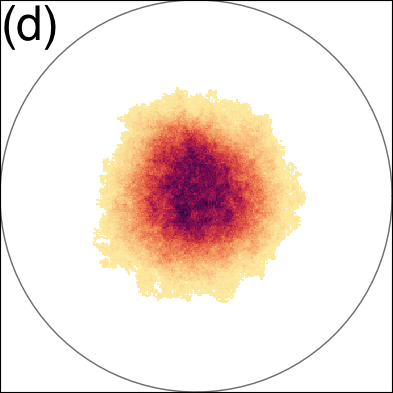} \includegraphics[width=0.19\linewidth]{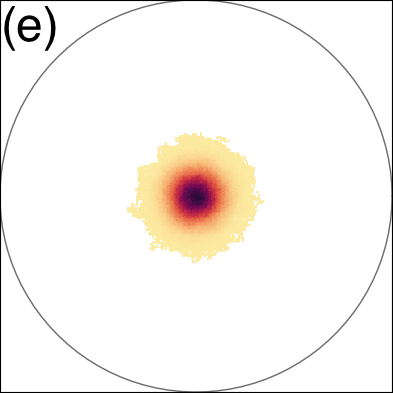} \\
\includegraphics[width=0.19\linewidth]{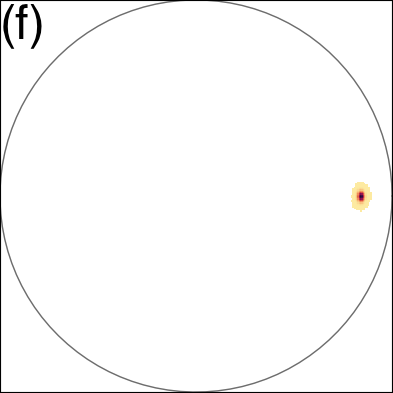} \includegraphics[width=0.19\linewidth]{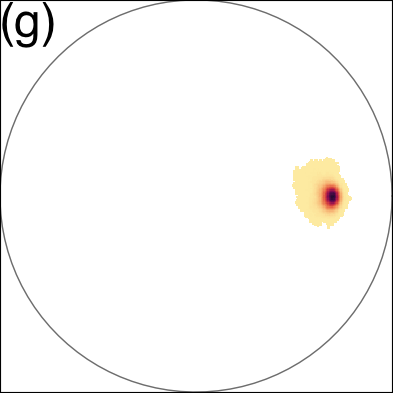} \includegraphics[width=0.19\linewidth]{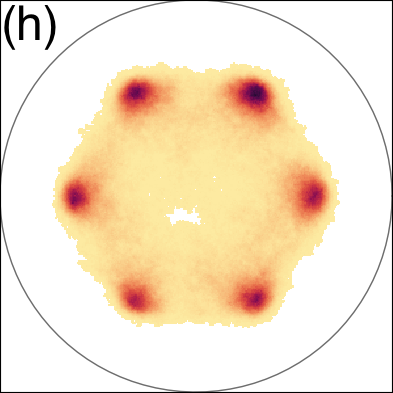} \includegraphics[width=0.19\linewidth]{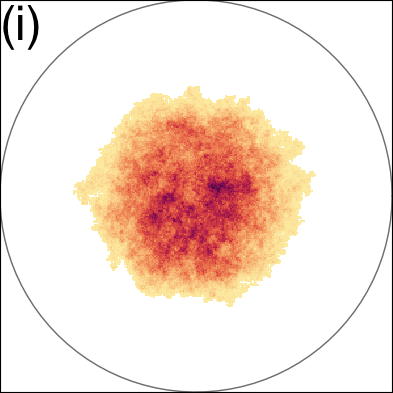} \includegraphics[width=0.19\linewidth]{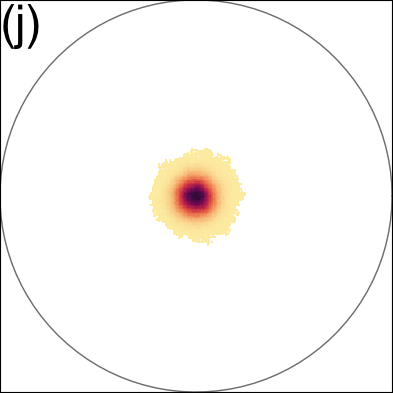}
\caption{Thermal order parameter $\Psi$ distributions for the clean $46 \times 46$ system in the in-plane phase at temperatures (a) $T = 1.0$, (b) $T = 1.5$, (c) the critical temperature $T_c = 1.56$, (d) $T = 1.7$, and (e) $T = 2.0$, and for the clean $64 \times 64$ system in the out-of-plane phase at temperatures (f) $T = 1.0$, (g) $T = 1.4$, (h) the critical temperature $T_c = 1.49$, (i) $T = 1.6$, and (j) $T = 2.0$. The circle in each plot represents the unit circle of the complex plane.\label{PsiSamples}}
\end{figure*}

The heat capacity and order parameter $\Psi$ for each system are plotted as a function of temperature in Figs.~\ref{CleanCs} and \ref{CleanOPs}. Both phases show a single sharp transition at which the heat capacity appears to be diverging with system size. Fig.~\ref{CleanOPs} shows remarkably large finite-size effects; even for very large systems of $16,384$ spins, the ordering transitions are not sharp enough to allow us to identify the critical temperatures by inspection. We extracted the critical temperatures by plotting the Binder ratios $Q := \langle |\Psi|^2 \rangle / \langle | \Psi | \rangle^2$ as a function of temperature for several system sizes. These ratios are constructed to be scale-independent at the critical temperature \cite{Binder}, and we found that the different size systems' curves crossed at the critical temperatures
\[
T_c = 1.5600 \pm 0.0025
\]
for the system in the in-plane phase and
\[
T_c = 1.4900 \pm 0.0025
\]
for the system in the out-of-plane phase.  In Fig.~\ref{PsiSamples} we plot the thermal distributions for the order parameter at temperatures below and above the transitions.

\begin{figure*}
\includegraphics[width=0.47\linewidth]{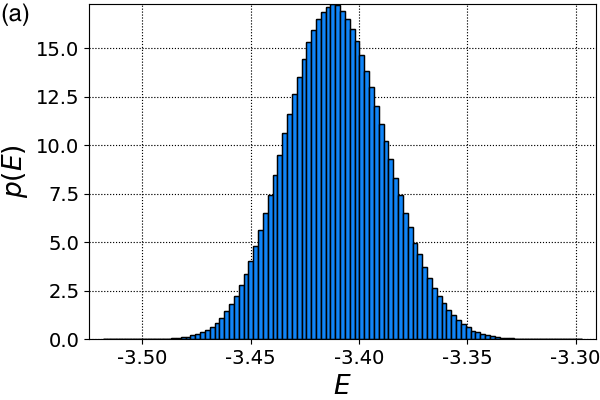} \hspace{10pt} \includegraphics[width=0.47\linewidth]{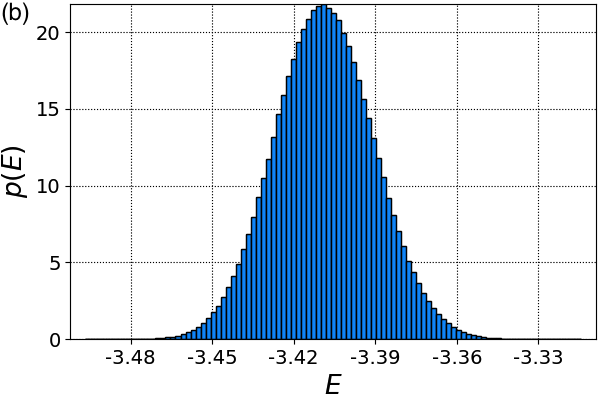}
\caption{Thermal energy distributions for the clean $128 \times 128$ systems in (a) the in-plane and (b) the out-of-plane phase. The distributions are single-peaked as expected for a continuous transition, rather than double-peaked as expected for a first-order transition. \label{Histograms}}
\end{figure*}

Both thermal energy distributions (plotted in Fig.~\ref{Histograms}) are single- rather than double-peaked at the transitions, indicating that the transitions are continuous. Moreover, the heat capacities diverge far more slowly than the volume-law divergence expected at a first-order transition. (Each system size plotted in Fig.~\ref{CleanCs} approximately doubles the previous plotted size, so if the transition were first-order, then we would expect the height of each curve's peak to double the height of the previous curve's peak.)

\begin{figure*}
\includegraphics[width=0.5\linewidth]{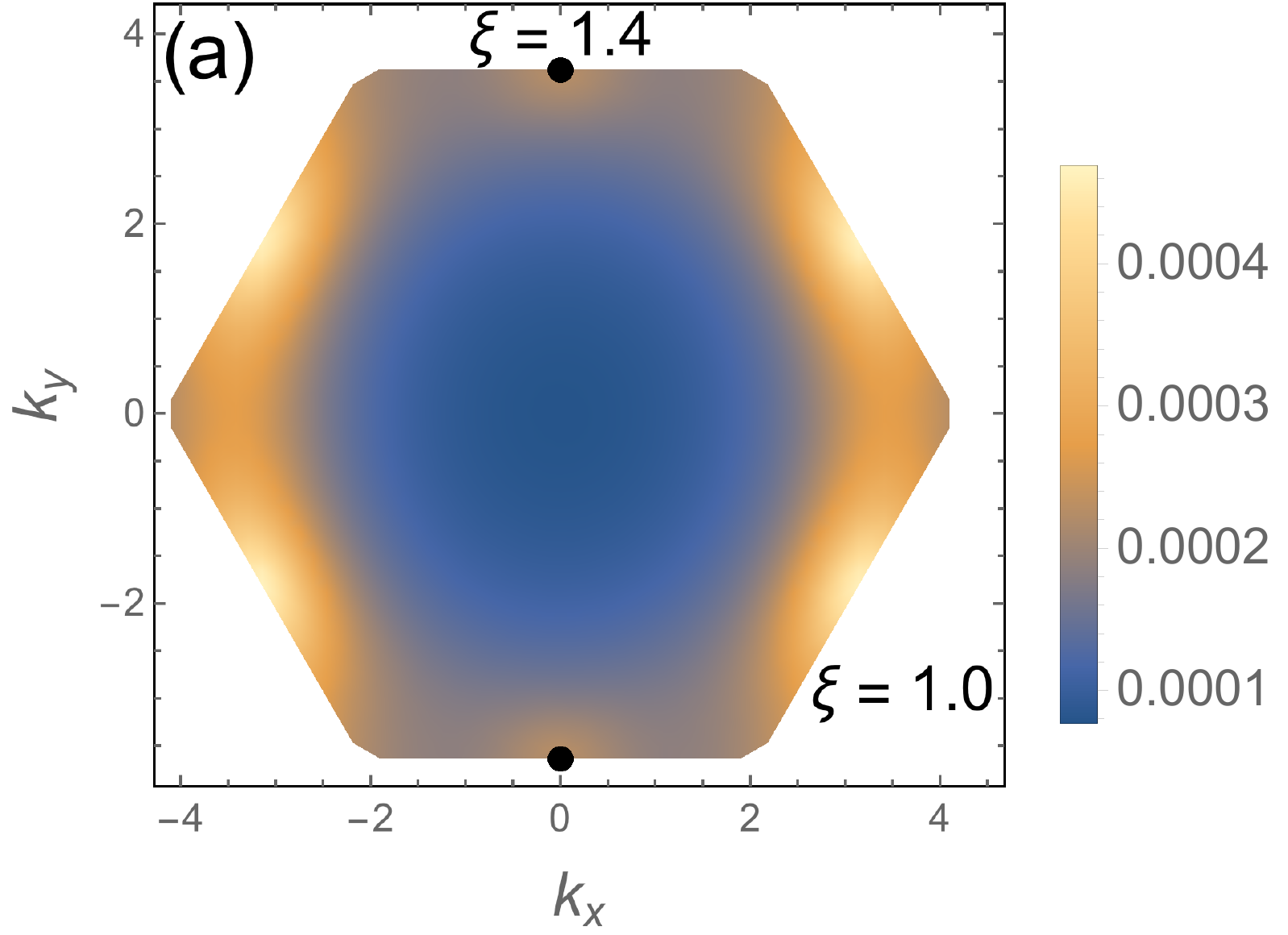} \hspace{2pt} \includegraphics[width=0.47\linewidth]{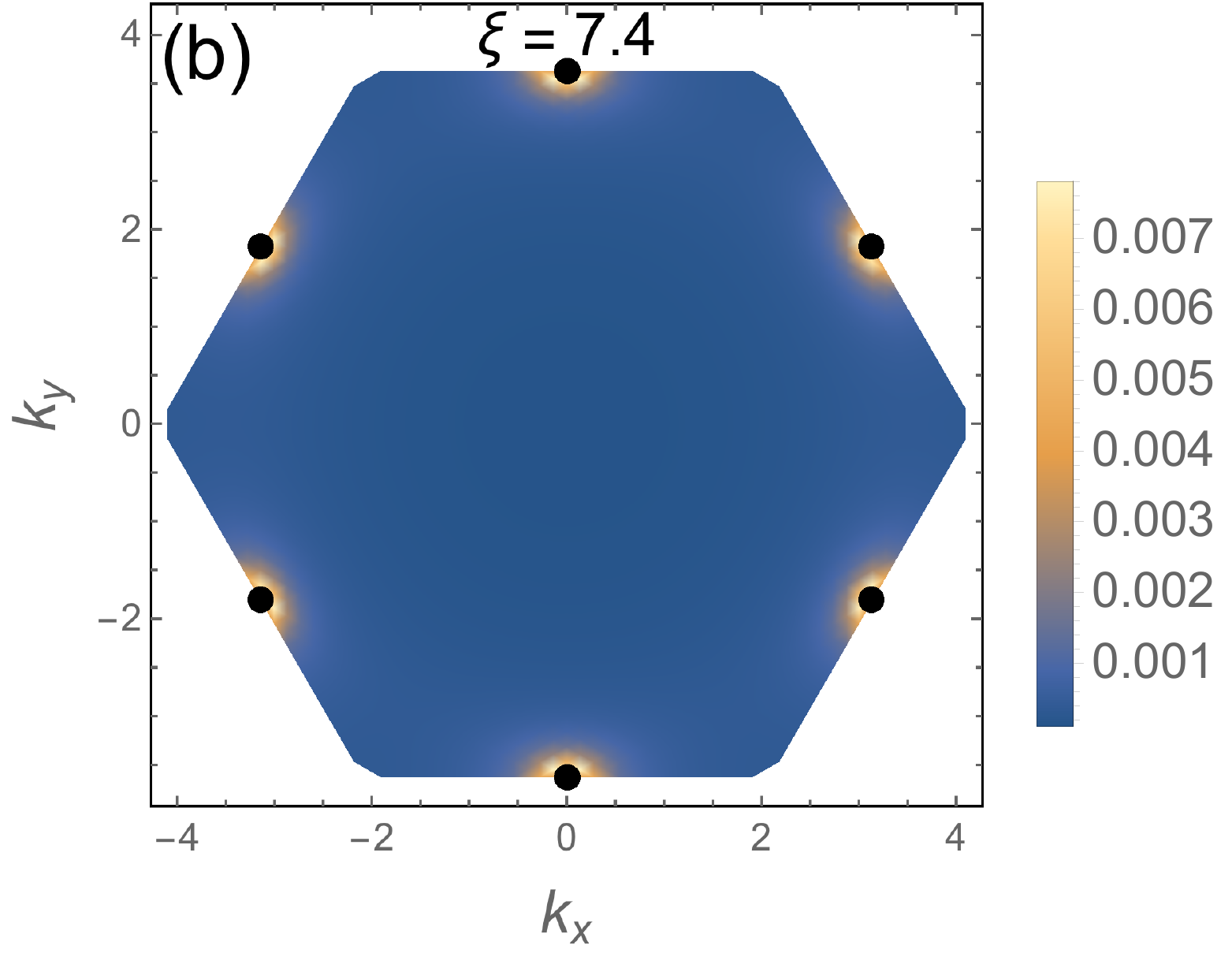} \\
\includegraphics[width=0.5\linewidth]{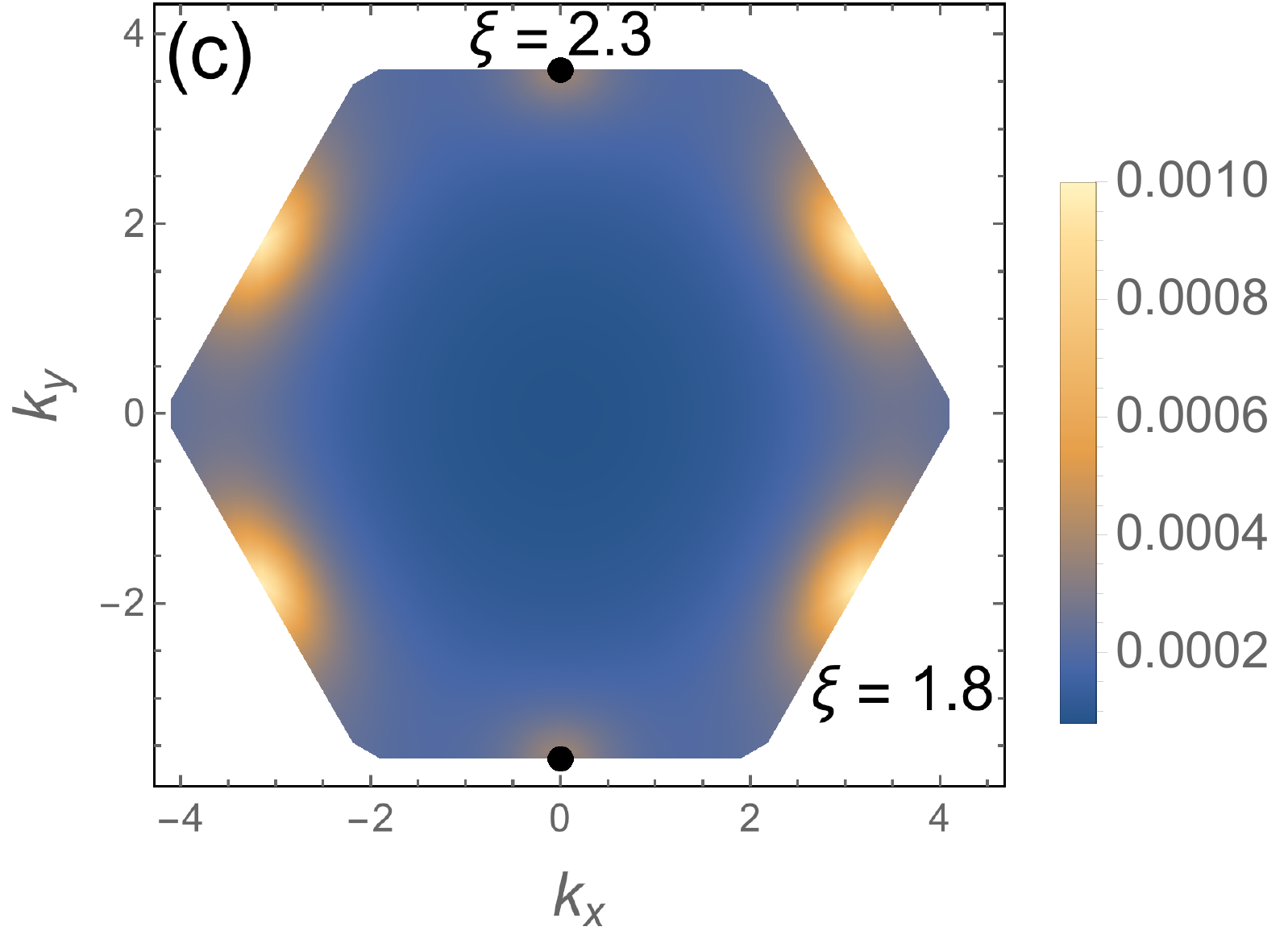} \hspace{2pt} \includegraphics[width=0.47\linewidth]{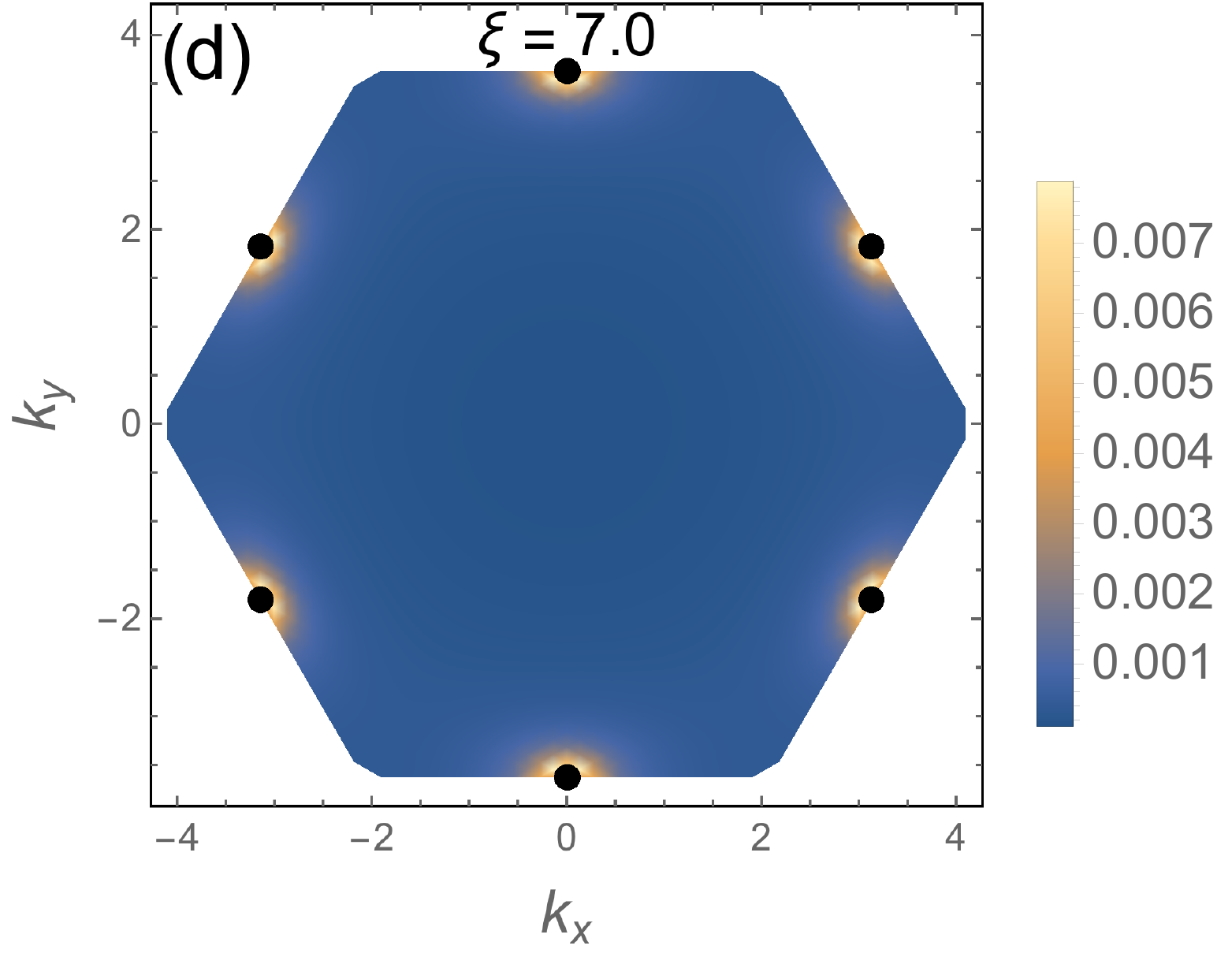}
\caption{Spin structure factors $S(\bm{k})$ defined by \eqref{Sk}, and connected correlation lengths in lattice spacings, for the clean $46 \times 46$ systems in the in-plane phase ((a)-(b)) and the out-of-plane phase ((c)-(d)), for temperatures $T = 1.3$ ((a), (c)) and $T = 1.7$ ((b), (d)). The spin structure factors at the critical temperatures look qualitatively similar to those for $T = 1.7$, except that a much higher fraction of the total spectral weight lies exactly at the $M$ points. The spectral weight at the $M$ points indicated by black dots is far higher than the plotted scale: they contain $62\%$ and $56\%$ of the total spectral weight for temperature $T = 1.3$, $32\%$ and $34\%$ at the critical temperatures (not shown), and $7\%$ for $T = 1.7$.} \label{Ss}
\end{figure*}

\subsubsection{Spin structure factor}

We also simulated the spin structure factor
\begin{align}
S(\bm{k}) &:= \frac{1}{N} \sum_{\bm{\Delta r}} C(\bm{\Delta r}) e^{-i \bm{k} \cdot \bm{\Delta r}}, \label{Sk} \\
C(\bm{\Delta r}) &:= \frac{1}{N} \sum_{\bm{r}} \langle \bm{S}_{\bm{r}} \cdot \bm{S}_{\bm{r} + \bm{\Delta r}} \rangle, \nn
\end{align}
which can be compared with the results of scattering experiments such as integrated inelastic neutron scattering. The structure factors are shown in Fig.~\ref{Ss} for temperatures below and above the critical temperature. In each subfigure, we indicate by black dots (shown to scale) the endpoints of individual reciprocal lattice vectors that contain a substantial fraction of the total spectral weight, which always lie exactly at the $M$ points of the Brillouin zone. We plot the remaining spectral weight as a heatmap. Below the critical temperatures, one particular symmetry-breaking state is spontaneously selected during the equilibration stage, and the resulting long-range order at one particular $M$ point is reflected in the high spectral weight indicated by the black dots.

We can estimate the connected correlation lengths $\xi$ by taking cuts along the edges of the Brillouin zone and fitting the resulting curves to Lorentzian distributions
\[
S(\bm{k}_M + \bm{q}) \propto \frac{1}{1 + (\xi q)^2},
\]
where $\bm{k}_M$ represents that edge's $M$ point. (Below the critical temperature, we neglect the single reciprocal lattice vector indicated by the black dots, as the spectral weight at this vector reflects the long-range order and we want the connected correlation length.) We find excellent fits near the $M$ points for temperatures above the critical temperature, and also below the critical temperature if we include a constant offset, representing a uniform background arising from nonuniversal short-distance physics. The fitted values of $\xi$ for inequivalent $M$ points are shown in the figure. Even fairly high above the critical temperature, the correlation lengths are a significant number of lattice spacings, explaining the large finite-size effects displayed in Fig.~\ref{CleanOPs} and the difficulty of reaching the thermodynamic limit.

\subsection{Theoretical analysis of ordering transitions}

As discussed above, the slow divergence of the heat capacity and the single-peaked energy distribution at the transitions indicate that both systems' transitions are continuous and should therefore show universal behavior. In this subsection we present two theoretical analyses of the ordering behavior -- one based on a comparison with previously studied $\mathbb{Z}_6$-symmetric lattice models and one based on field theory.

\subsubsection{Comparison with Potts-type models}

We can compare our ordering transitions with the known transitions in previously studied 2D nearest-neighbor classical models whose ground state manifolds also have $\mathbb{Z}_6$ symmetry. The models that have been studied in by far the most depth are the spatially isotropic Potts-type models, in which each spin can take on six possible configurations, with the nearest-neighbor couplings respecting a $\mathbb{Z}_6$ symmetry but otherwise arbitrary. These models have a very rich phase diagram \cite{Cardy80, AlcarazKoberle80, AlcarazKoberle81}, with three (non-multicritical) types of transitions:
\begin{enumerate}
\item A single first-order $\mathbb{Z}_6$ breaking transition \cite{Baxter, Wu} 
\item A Kosterlitz-Thouless (KT) transition at an upper critical temperature and an \emph{inverted} KT transition (which spontaneously breaks the $\mathbb{Z}_6$ symmetry) at a lower critical temperature, surrounding a ``massless'' phase in which the correlation functions are power-law with an exponent that varies continuously with temperature \cite{Jose, Challa, Lapilli, PricePerkins12, PricePerkins13}
\item Separate second-order $\mathbb{Z}_2$- and $\mathbb{Z}_3$-symmetry breaking at two different temperatures (in either order) \cite{PricePerkins13}.
\end{enumerate}
The two KT transitions and the first-order transition all meet at a multicritical ``Fateev-Zamolodchikov'' (FZ) line at which the model is integrable \cite{FZ}. This entire line has a single renormalization-group (RG) IR fixed point, which is described by the same minimal-model conformal field theory (CFT) as the antiferromagnetic RSOS model \cite{ZF, Alcaraz, Bonnier}.

Unfortunately, there are at least three reasons why these models probably do not make a good analogy for our systems. First, our systems' ground-state manifolds have $\mathbb{Z}_6$ symmetry, but it is not clear that the finite-temperature states below the ordering transitions do as well -- the structure of the ground states suggests that thermal fluctuations might reduce the symmetry to $\mathbb{Z}_3 \times \mathbb{Z}_2$. Second, as we explain below, we believe that the spin-orbit coupling, which is not captured by the Potts-type models, is qualitatively important at the transitions (even beyond the fact that it explicitly breaks the symmetry group down to a discrete group and allows finite-temperature ordering).

\begin{figure*}
\includegraphics[width=0.47\linewidth]{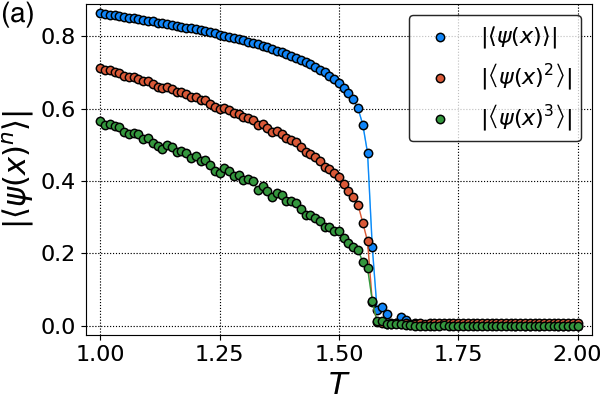} \hspace{10pt} \includegraphics[width=0.47\linewidth]{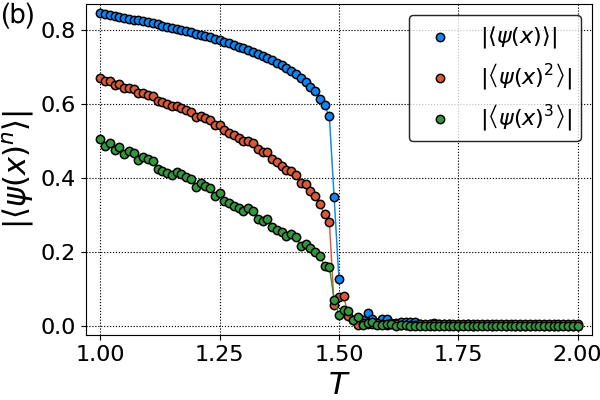}
\caption{Order parameters for $\mathbb{Z}_6$, $\mathbb{Z}_3$, and $\mathbb{Z}_2$ symmetry breaking for $128 \times 128$ systems in the (a) in-plane and (b) out-of-plane phases. All three symmetry subgroups appear to break together. The angle brackets denote a spatial as well as thermal average. The differences between these curves and the corresponding curves for the smaller systems are smaller than the plot markers. \label{Psins}}
\end{figure*}

Third, our Monte Carlo results described in Sec.~\ref{Thermo} do not seem compatible with any of the Potts-model phase transitions described above:
\begin{enumerate}
\item The slow divergence of the heat capacity and the single-peaked energy distribution at the transitions rule out first-order transitions.
\item But the heat capacities do certainly appear to diverge with system size (albeit slowly), and at a double Kosterlitz-Thouless transition the heat capacity remains completely smooth in the thermodynamic limit (with an exponentially flat non-analytic contribution) \cite{Kosterlitz}. In particular, our Fig.~\ref{CleanCs} shows a \emph{much} sharper peak than Fig.~6 of Ref.~\onlinecite{PricePerkins12}, in which the heat capacity near an inverted KT transition shows a negligible dependence on system size.  Moreover, a continuous $\text{U}(1)$ symmetry emerges at an inverted KT transition \cite{Jose, Ortiz, Lapilli, PricePerkins13}, which would be indicated by a continuous ring of order parameter concentration with nonzero radius, like the one shown in Fig.~2(g) of Ref.~\onlinecite{PricePerkins12} -- but in our systems the discrete sixfold symmetry is still very clear at the critical temperatures in Fig.~\ref{PsiSamples}.
\item The heat capacities plotted in Fig.~\ref{CleanCs} only show a single peak, rather than the two separate peaks we would expect for separate $\mathbb{Z}_2$ and $\mathbb{Z}_3$ symmetry breaking. Moreover, the order parameter distributions plotted in Fig.~\ref{PsiSamples} do not clearly indicate that either symmetry subgroup is preferentially preserved at the critical temperature. Finally, the natural order parameters for these partial symmetry breakings are
\beq
\frac{1}{N / 4} \sum_x \left \langle \psi(x)^3 \right \rangle \quad \text{and} \quad \frac{1}{N / 4} \sum_x \left \langle \psi(x)^2 \right \rangle \label{psin}
\eeq
for the upper $\mathbb{Z}_2$ or $\mathbb{Z}_3$ symmetry breaking, respectively, with $\psi(x)$ defined in \eqref{psi}.  The former order parameter transforms nontrivially under spatial inversion $\mathcal{I}$ (which generates the $\mathbb{Z}_2$ symmetry subgroup and takes $\psi(x) \to -\psi(x)$) but trivially under the plane rotation $\mathcal{C}_3$ (which generates the $\mathbb{Z}_3$ symmetry subgroup and takes $\psi(x) \to e^{2 \pi i / 3} \psi(x)$), and the latter vice-versa. As shown in Fig.~\ref{Psins}, both symmetries appear to be broken at the same temperature (to within our simulations' temperature resolution) as the $\mathbb{Z}_6$ symmetry, providing further support for the existence of a single sixfold-symmetry-breaking transition for each system. (Note that in Fig.~\ref{CleanOPs} we plot $\langle | \psi(x) | \rangle$ but in Fig.~\ref{Psins} we plot $|\langle \psi(x) \rangle|$. Calculating $|\langle \psi(x) \rangle|$ effectively combines the thermal and spatial averages into a single ``spacetime'' average, which greatly increases the effective system size and sharpens the transitions, but a systematic finite-size scaling analysis of $|\langle \psi \rangle|$ is more difficult than of $\langle |\psi| \rangle$.)
\end{enumerate}

\begin{figure*}
\includegraphics[width=0.49\linewidth]{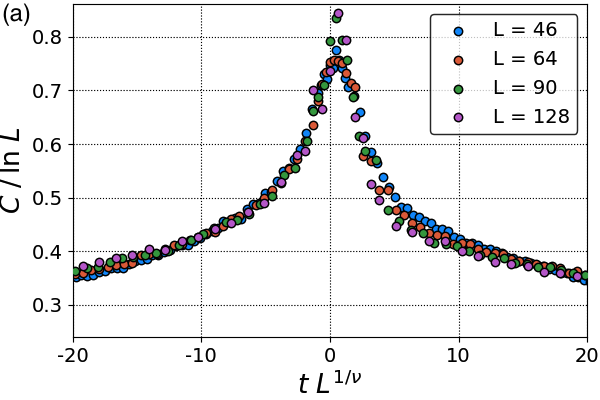} \hspace{2pt} \includegraphics[width=0.49\linewidth]{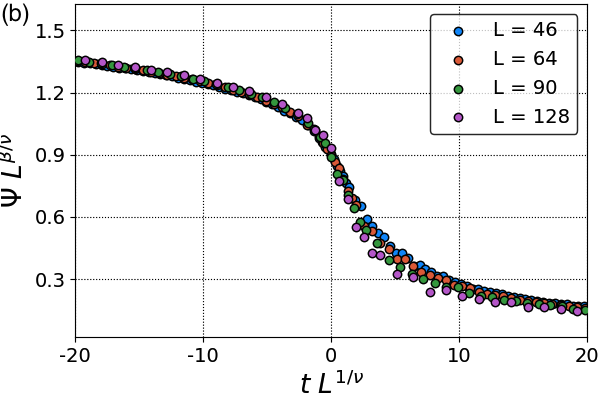} \\
\includegraphics[width=0.49\linewidth]{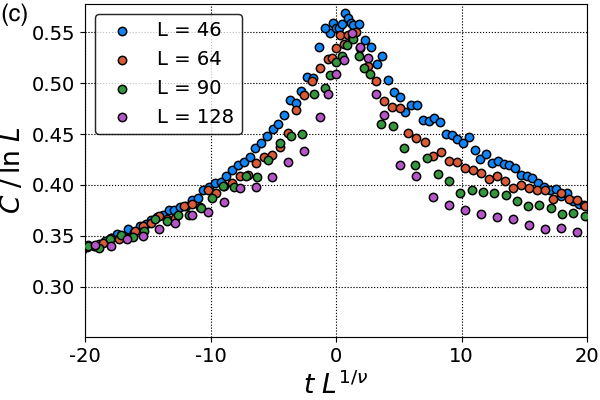} \hspace{2pt} \includegraphics[width=0.49\linewidth]{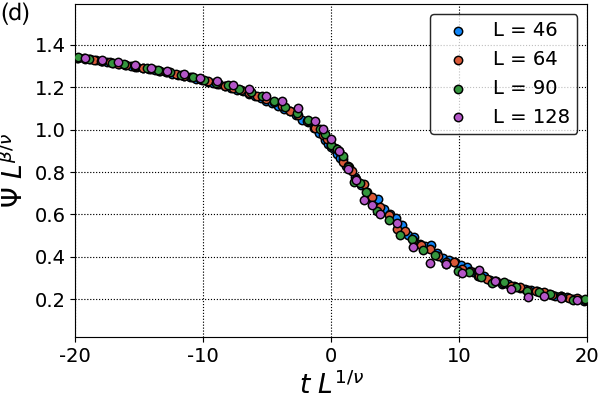} \\
\caption{Heat capacity and order parameter collapses for the Ising critical exponents $\beta = 1/8,\ \nu = 1$ via \eqref{LogScaling} and \eqref{PsiScaling} for the clean system in the (a)-(b) in-plane and (c)-(d) out-of-plane phase.} \label{Z2Collapses}
\end{figure*}

\begin{figure*}
\includegraphics[width=0.49\linewidth]{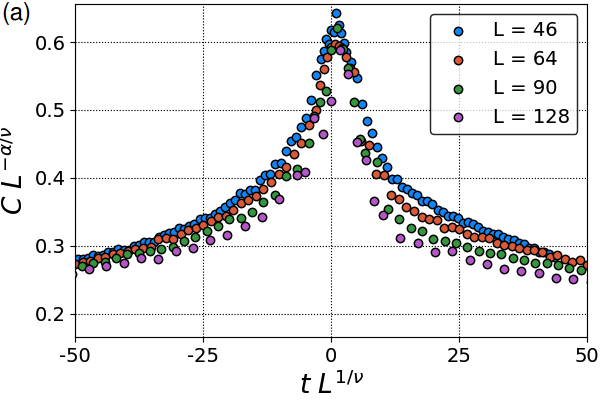} \hspace{2pt} \includegraphics[width=0.49\linewidth]{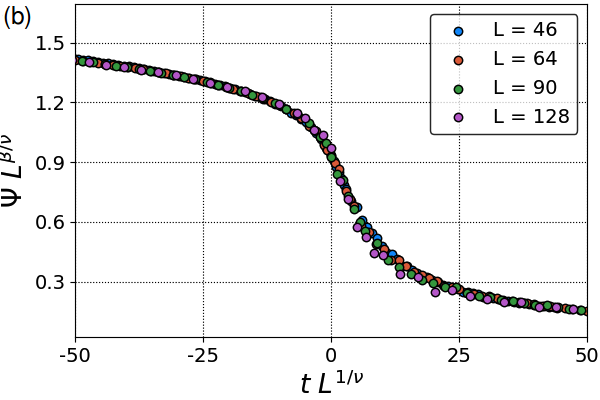} \\
\includegraphics[width=0.49\linewidth]{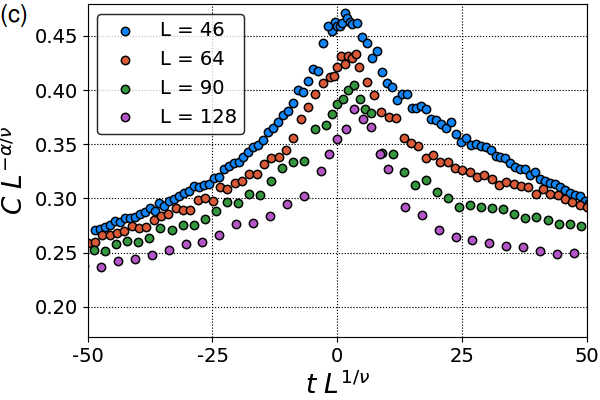} \hspace{2pt} \includegraphics[width=0.49\linewidth]{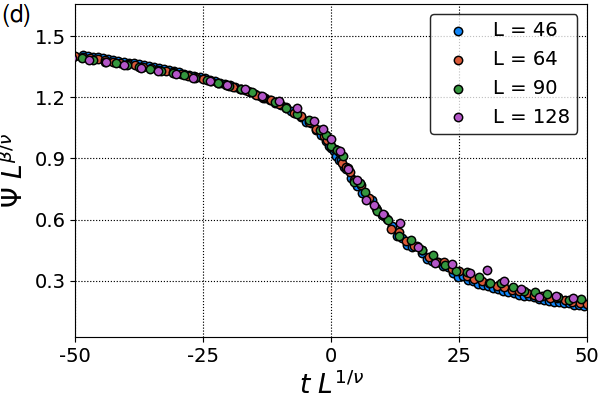}
\caption{Heat capacity and order parameter collapses for the $\mathbb{Z}_3$ universality class critical exponents $\alpha = 1/3,\ \beta = 1/9,\ \nu = 5/6$ via \eqref{CScaling} and \eqref{PsiScaling} for the clean system in the (a)-(b) in-plane and (c)-(d) out-of-plane phase.} \label{Z3Collapses}
\end{figure*}

We also performed a finite-size scaling analysis on our data by attempting to collapse the heat capacity and order parameter curves to universal functions via the usual scaling equations
\begin{align}
C L^{-\alpha / \nu} &= f_C \left( t L^{1 / \nu} \right), \label{CScaling} \\
\Psi L^{\beta / \nu} &= f_\Psi \left( t L^{1 / \nu} \right), \label{PsiScaling}
\end{align}
where $L$ is the linear size of the system and $t := (T - T_c)/T_c$ is the reduced temperature.  If the $\mathbb{Z}_2$ and $\mathbb{Z}_3$ symmetry subgroups were broken at separate temperatures, then we would expect the lower transition's critical exponents to be either the Ising critical exponents $\alpha = 0$, $\beta = 1/8$, $\nu = 1$, or the $\mathbb{Z}_3$ universality class critical exponents $\alpha = 1/3,\ \beta = 1/9,\ \nu = 5/6$ \cite{Wu}. (In the Ising case, \eqref{CScaling} is replaced by
\beq
\frac{C}{\ln L} = f_C \left( t L^{1/\nu} \right).) \label{LogScaling}
\eeq
The putative scaling collapses for these critical exponents are plotted in Figs.~\ref{Z2Collapses} and \ref{Z3Collapses}. The order-parameter data collapse is reasonably good for both universality classes. The heat-capacity data collapse is fair for the in-plane system and the Ising universality class (except very close to the transition, where critical slowing down makes equilibration difficult), but for the out-of-plane system it is not very good for either universality class.

The transitions are very unlikely to be in the FZ universality class, as the FZ transition is multicritical and our couplings are not in any way fine-tuned. Nevertheless, we checked the data scaling collapse for the FZ critical exponents $\alpha = \nu = 2/3$ \cite{Alcaraz, Bonnier}, and found much worse agreement than for either the Ising or $\mathbb{Z}_3$ critical exponents, as our systems' heat capacities diverge much more slowly than predicted by the fairly large FZ critical exponent $\alpha = 2/3$.

\subsubsection{Field-theory analysis}

The free energy has a minimum at each of the three inequivalent $M$ points of the Brillouin zone (App.~\ref{GS app}), so we need only consider the vicinity of these points and can consider three slowly varying order-parameter fields $\varphi_b(\bm{q}) := \varphi \left( \bm{k}_M^{(b)} + \bm{q} \right)$, where the valley index $b$ labels the $M$ point as in Sec.~\ref{OP} and $|\bm{q}|$ is much less than the inverse lattice spacing. A key point (which the previous subsection's models fail to reflect) is that even though the spins are Heisenberg spins, the spin-orbit coupling explicitly picks out a favored (unoriented) direction $\pm \bm{n}_b$, where $\bm{n}_b$ is given by \eqref{IPn} and \eqref{OPn}. Since we are now considering each $M$ point individually, it is natural to replace the microscopic order parameter \eqref{Psi} by the position-dependent version of \eqref{SdotSb}:
\[
\varphi_b(\bm{x}) = \bm{S}_{\bm{x}} \cdot e^{i \bm{q}_b \cdot \bm{x}} \bm{n}_b = (-1)^{\Delta_b(\bm{x})} \bm{S}_{\bm{x}} \cdot \bm{n}_b,
\]
which is a real scalar field that takes on the value $(-1)^p = \pm 1$ in the two ground states corresponding to the $M$ point $b$.

These coarse-grained fields have the Landau free energy density
\begin{align}
\mathcal{F} =\, &\sum_{b = 1}^3 \left[ \frac{1}{2} (\bm{\del} \varphi_b)^2 + \frac{1}{2} r\, \varphi_b^2 + \frac{1}{4!} u\, \varphi_b^4 \right] \label{F} \\
&+ \frac{1}{4} v\, \sum_{b, b' = 1}^3 \varphi_b^2 \varphi_{b'}^2 + \dots, \nn
\end{align}
with all scalar interactions marginal in 2D. (We have used the Hamiltonian's $\mathcal{C}_3$ and time-reversal symmetries, which take $\varphi_b$ to $\varphi_{b+1}$ and to $-\varphi_b$ respectively, to eliminate terms odd in $\varphi_b$ and to reduce the number of independent couplings.) If we collect the three scalar fields $\varphi_b$ into an abstract Euclidean $3$-vector $\vec{\varphi}$ in ``valley space,'' we can rewrite the free energy density as
\begin{align*}
\mathcal{F} =& \, \frac{1}{2} (\bm{\del} \vec{\varphi})^2 + \frac{1}{2} r\, \vec{\varphi}^2 + \frac{1}{4!} u\, (\vec{\varphi}^2)^2 \\
&+ \frac{1}{4} \left( v - \frac{1}{3} u \right) \sum_{b, b' = 1}^3 \varphi_b^2 \varphi_{b'}^2 + \dots.
\end{align*}
If $v = u / 3$ (and the higher couplings are tuned similarly), then the theory's $\mathcal{S}_3 \times (\mathbb{Z}_2)^3$ symmetry is enlarged to a continuous $\text{O}(3)$ symmetry. Our numerical simulations indicate that only one component $\varphi_b$ acquires a nonzero expectation value as we tune $r \propto T - T_c$ below zero, and that this transition is continuous. These two results indicate that $v > u/3$ and $u > 0$ respectively.

The free energy \eqref{F} describes three coupled Ising models. We would like to know whether at the transition, the couplings between the Ising models flow to weak or to strong coupling under RG, which would affect the transition's universality class. Unfortunately, this question is difficult to study analytically, as there are infinitely many marginal couplings and the IR fixed point of the 2D scalar field theory describing a single critical Ising model is already strongly coupled. (The strong coupling between the three Ising models with coupling constant $v > u/3$ prevents us from using the standard mapping from a massless 2D real scalar field to a massless free Majorana-fermion field, which would map the action \eqref{F} to a much more analytically tractable $\mathrm{SO}(3)_1$ Wess-Zumino-Witten model \cite{DiFrancesco}).

Ref.~\onlinecite{Nienhuis} studied a discretized version of \eqref{F} and concluded that the transition between the symmetric and fully ordered phases is always first-order. The authors' choice of discretization effectively only captured $\varphi^4$ interactions; it is possible (though unlikely) that higher-order interactions could be marginally relevant under RG and change the nature of the phase diagram. Unfortunately, this possibility is difficult to study analytically, as it would dependent sensitively on the higher-order coupling constants and therefore on the microscopic details of the model. Putting aside this subtlety, the field-theory analysis seems to predict a first-order transition which is difficult to square with the numerical results described above. The transitions' universality class therefore remains an open question.

\section{Finite-temperature behavior of disordered systems \label{Disordered MC}}

Experimental samples of \Yb\ are believed to have significant random mixing of the nonmagnetic $\text{Mg}^{2+}$ and $\text{Ga}^{3+}$ ions, which may strongly effect the effective pseudospin $g$ factors, and potentially the magnetic couplings as well \cite{YLi17a}. Ref.~\onlinecite{Zhu17} used DMRG on small clusters to study a disorder pattern in a model similar to \eqref{Ham} in which the signs of the (fixed-magnitude) $J_{\pm\pm}$ couplings are chosen randomly on each bond. The authors found that in certain regimes, the ground state of the disordered model breaks into essentially classical domains with different stripe orientations, while in other regimes, it forms domains which are collective superpositions of two different stripe-ordered states. While they were unable to reach large enough system sizes to conclusively determine whether single domains could grow extensively, they proposed that the disordered model's ground state lacks long-range order.

We take a similar line of approach, but with a less drastic disorder model. Rather than flipping the sign of any couplings, we chose a model in which we randomly and independently multiply each bond strength by a number uniformly distributed on the interval $[1 - \Delta, 1 + \Delta]$ for some nonnegative parameter $\Delta$ that quantifies the disorder strength. The case $\Delta = 0$ corresponds to no disorder, and if $\Delta \leq 1$ (the only regime that we consider), then none of the bonds change between being anti- and ferromagnetic. Like Ref.~\onlinecite{Zhu17}, we do not present any precise physical argument to support this model, but we believe that the results should not depend sensitively on the details of the disorder -- however, we emphasize the important point that our disorder model does not introduce any random fields, and so preserves exact time-reversal symmetry (like Ref.~\onlinecite{Zhu17}'s model). The effect of bond disorder on the standard Potts model has been studied extensively via renormalization-group methods and numerics \cite{Cardy97}, but to our knowledge, bond disorder in spin-orbit-coupled models has only come under theoretical study very recently \cite{Andreanov, Ross, Zhu17, Andrade, Kimchi}. As we discuss below, we expect the effects of disorder to manifest themselves very differently in the SO case.

\begin{figure*}
\includegraphics[width=0.49\linewidth]{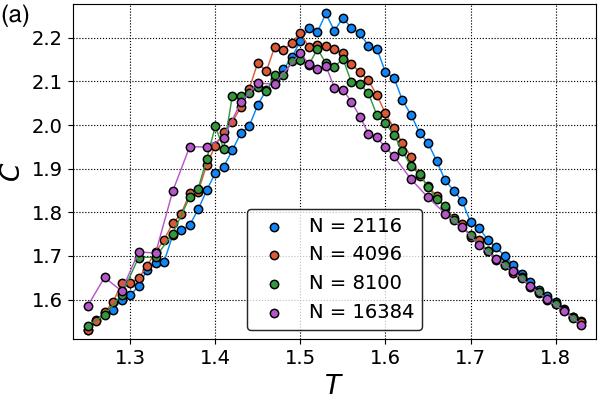} \hspace{2pt} \includegraphics[width=0.49\linewidth]{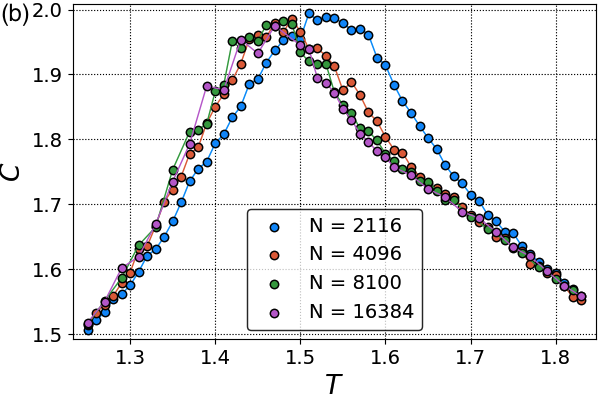} \\
\caption{Disorder-averaged heat capacity per site for (a) the in-plane-phase system at disorder strength $\Delta = 0.3$ and (b) the out-of-plane-phase system at disorder strength $\Delta = 0.2$. \label{DisorderedCs}}
\end{figure*}

Fig.~\ref{DisorderedCs} shows the heat capacity in the presence of fairly weak disorder. Each system is averaged over 10 disorder realizations. (The systems are large enough and the disorder weak enough that the disorder self-averages well, and converged thermodynamic quantities show very little variation between realizations. But even the light disorder significantly increases the thermal equilibration times, so the thermal rather than the disorder averaging is the numerical bottleneck that makes the curves in Fig.~\ref{DisorderedCs} less smooth than the clean systems' in Fig.~\ref{CleanCs}. The parallel-tempering Monte Carlo sampling algorithm is typically more effective in strongly disordered systems than the Metropolis algorithm that we used, and may be able to overcome these thermalization barriers; we leave this possibility for future work.) Finite-size effects are important over a much broader temperature range than in the clean case (e.g.\ the curves for the $46 \times 46$ systems are noticeably shifted to the right relative to the larger systems'), suggesting that even light disorder greatly increases the connected correlation length.

The diverging peak at the critical temperature has completely vanished, and the heat capacity curve appears to have reached a smooth thermodynamic limit, suggesting the absence of any phase transition at this disorder level. Measuring the order parameter in the disordered system is challenging because it takes an extremely large number of sweeps to equilibrate. For both systems, the order parameter is nonneglible at low temperatures at the system sizes that we were able to reach, but its magnitude decreases sharply with increasing system size and appears to vanish in the thermodynamic limit. Together, these results suggest that these disordered systems do not display long-range order at any positive temperature.

The curves remain similar for lower disorder strengths, with the peak in the heat capacity greatly diminished relative to the clean case. As we explain below, we conjecture that for any positive disorder strength, the heat capacity of very large systems will eventually saturate at a finite value and there will be no phase transition, although we were unable to reach that saturation size in the case of very weak disorder.

Part of the destruction of long-range order in the disordered system is straightforward to understand via a generalization of the argument given by Imry and Ma for the ground state of the classical random-field Ising model (RFIM) in $d$ dimensions \cite{Imry}. In this model, large domains of linear size $L$ have random net field imbalances that scale as $L^{d/2}$, so by locally ordering into a domain aligned with the nearby field imbalance, the system can gain an energy of $o(L^{d/2})$, at the cost of a domain wall whose energy scales as $L^{d-1}$. For $d > 2$ the domain wall cost dominates and the ground state remains long-range ordered, but for $d \leq 2$ the ground state fractures into multiple large domains for arbitrarily weak disorder. This argument was later made completely rigorous \cite{Aizenman}.

Ref.~\onlinecite{Aizenman} also demonstrates that the finite-temperature ordering transition of the two-dimensional classical random-\emph{bond} Potts model (RBPM) \emph{does} survive at weak disorder (although for $n > 4$, when the clean system's transition is first-order, the bond disorder ``rounds'' it to a second-order transition \cite{Cardy99}). The RBPM might initially seem to be a better analogy than the RFIM is for our bond-disordered model \eqref{Ham}, but the presence of spin-orbit coupling in our model actually makes it more closely analogous to the RFIM: the key point is that in both our model \eqref{Ham} and the RFIM, the disorder couples directly to the order parameter and explicitly breaks the Hamiltonian symmetry that is spontaneously broken in the clean case \cite{Aharony, Andrade}. This is not the case in the RBPM, where the bond disorder preserves the exact $\mathbb{Z}_n$ spin-space symmetry.

\begin{figure*}
\includegraphics[width=0.49\linewidth]{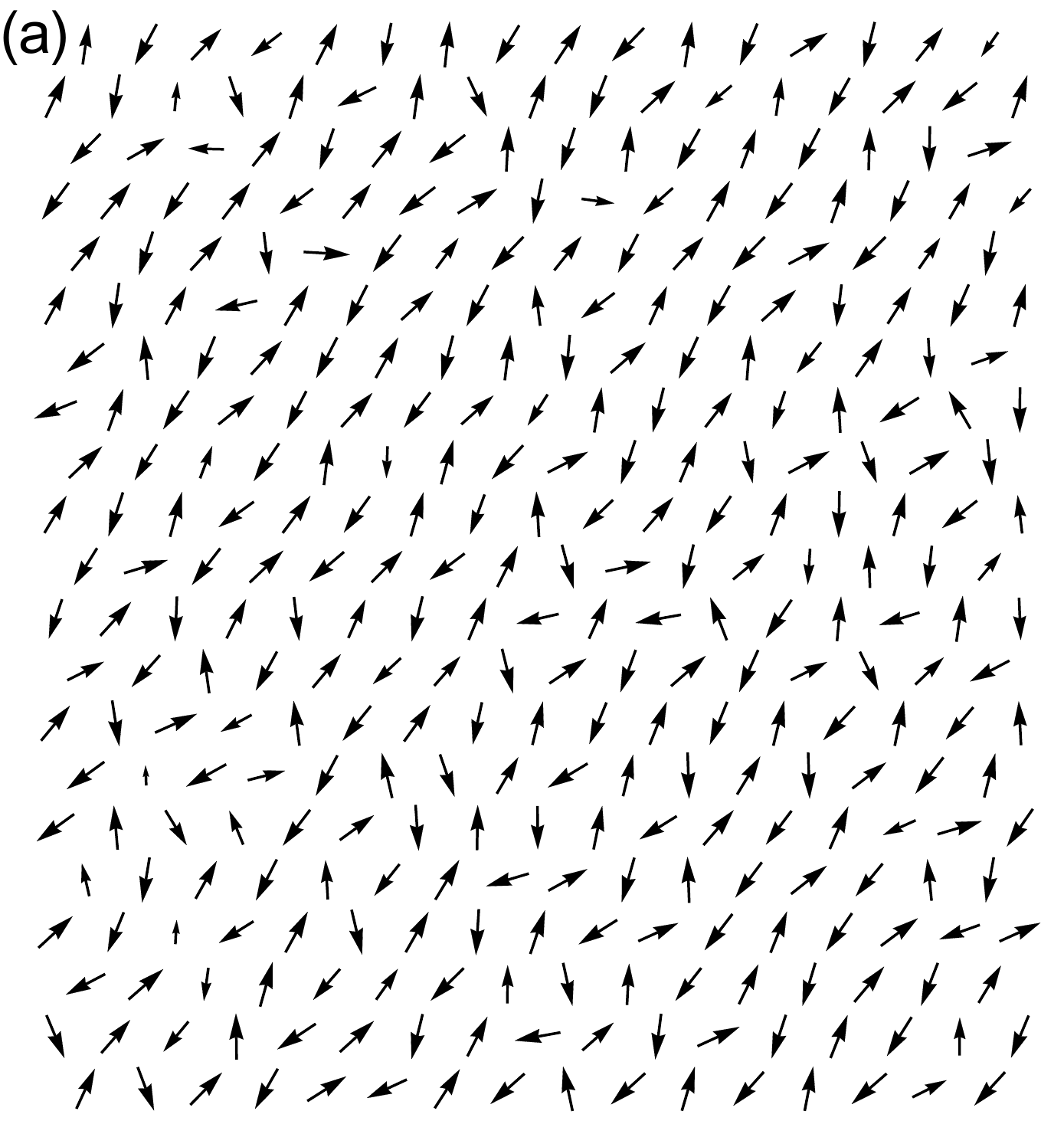} \hspace{2pt} \includegraphics[width=0.49\linewidth]{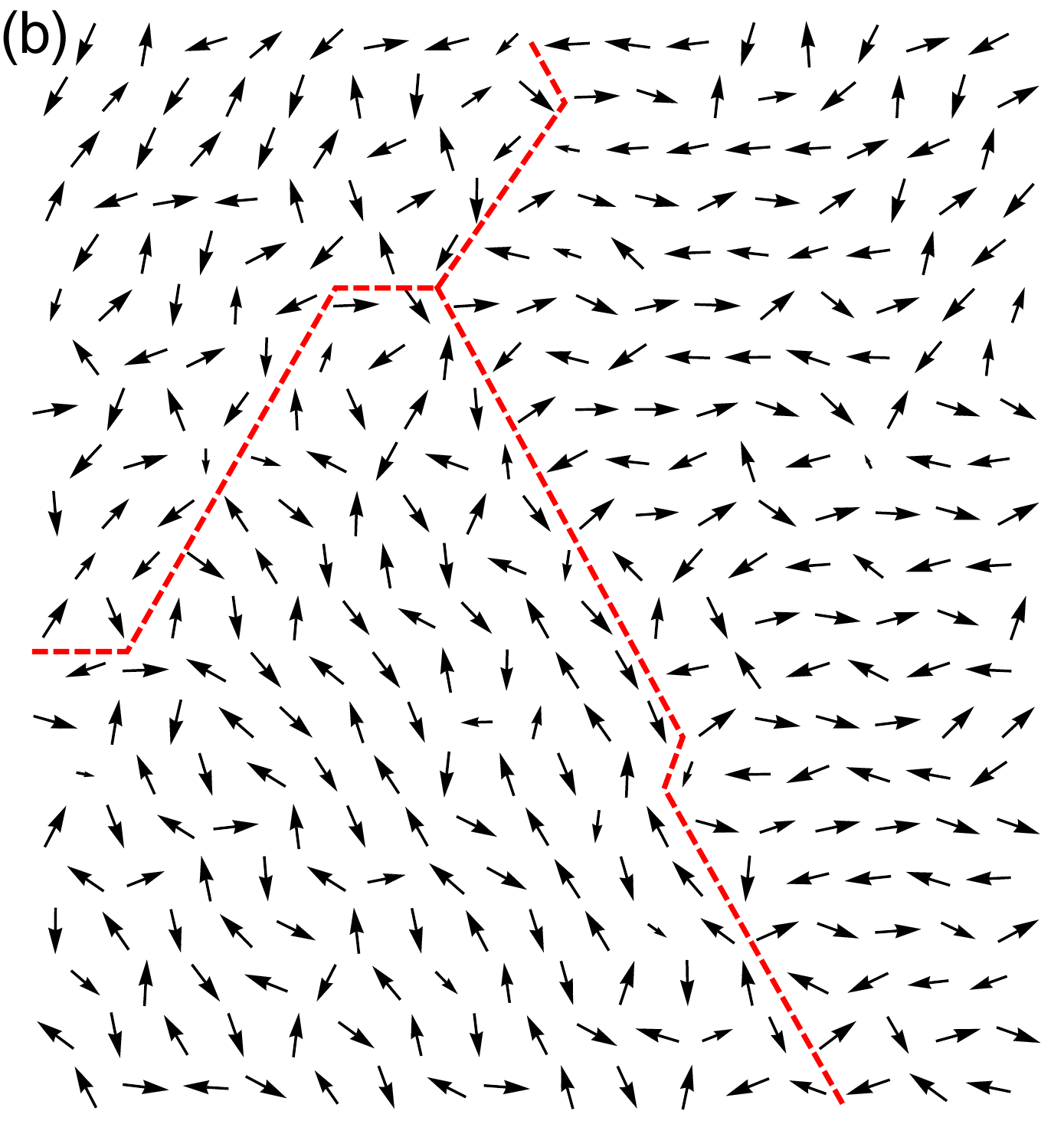}
\caption{Snapshots of part of the in-plane-phase system at temperature $T = 1.0$ in the case of (a) no disorder and (b) strong disorder $\Delta = 0.8$. (a) displays visible long-range order with stripes running parallel to the $\bm{a}_3$ direction. (b) shows the intersection of three domains whose stripes run parallel to the $\bm{a}_1$ (right), $\bm{a}_2$ (bottom left), and $\bm{a}_3$ directions (top left), with the domain walls indicated by red dashed lines. \label{Snaps}}
\end{figure*}

We therefore understand why our model probably cannot display long-range stripe order at any temperature. For any given disorder realization, there are large domains in which the bonds parallel to one of the $\bm{a}_1$, $\bm{a}_2$, or $\bm{a}_3$ directions happen to be significantly stronger than the bonds parallel to the other two directions. (In the terminology of Sec.~\ref{OP}, each domain has a preferred value of $b$ or color in Fig.~\ref{GSs}.) The system locally orders into stripes running along that domain's favored direction, but since different domains have different favored directions, long-range stripe order is not possible. Since these locally favored directions are determined by the quenched disorder realization, which explicitly breaks the Hamiltonian's symmetry, each domain's value of $b$ does not fluctuate over time, nor do the domain wall locations fluctuate very much. The disorder strength sets the scale of the domain sizes, and the system must contain many domains in order for the heat capacity to saturate and the order parameter to vanish. In Fig.~\ref{Snaps} we show representative snapshots of parts of the in-plane-phase system, with all three domain types clearly visible in the displayed region. Note that even at the quite strong disorder strength $\Delta = 0.8$, the domains are still very large, which explains why we need an appreciable disorder strength to observe the destruction of long-range order in our systems.

As mentioned above, our disorder model preserves exact time-reversal (TR) symmetry, just as the bond disorder in the RBIM preserves exact $\mathbb{Z}_n$ symmetry. A given disorder realization will create domains which favor an unoriented direction for the stripes to run, but no favored orientation of spins within each stripe. (In the terminology of Sec.~\ref{OP}, a given domain does \emph{not} have a preferred value of $p$.) The Imry-Ma argument therefore still allows for the possibility of a (potentially rounded) phase transition in which TR symmetry is spontaneously and globally broken, just as the RBPM still has a finite-temperature transition in the presence of bond disorder that does not couple directly to the order parameter. The simplest global TR-symmetry-breaking pattern would be one in which the system spontaneously chooses a single value of $p$ for \emph{all} the domains -- in other words, it locally aligns into either the three ground states indicated by circles in Fig.~\ref{GSs}, or the three indicated by squares. The local order parameter corresponding to this TR symmetry breaking would be the quantity $\left \langle \psi(x)^3 \right \rangle$ introduced in \eqref{psin} in the context of the clean system's ground state manifold's $\mathbb{Z}_2$ symmetry being broken at a strictly higher temperature than a $\mathbb{Z}_6 \to \mathbb{Z}_2$ transition, as cubing the order parameter removes the dependence on the domain type $b$.

Nevertheless, we see no sign of such a TR-symmetry breaking transition in our data. Fig.~\ref{DisorderedCs} does not show any sharp peak in the heat capacity, and the putative order parameter $\left \langle \psi(x)^3 \right \rangle$ appears to go to zero with increasing system size just as quickly as $\langle \Psi \rangle \equiv \langle \psi(x) \rangle$, which the Imry-Ma argument guarantees will vanish in the thermodynamic limit of a disordered system.

We believe that the fragmentation of the ground state into domains is the explanation for this lack of homogeneous TR-symmetry breaking. In the case of the weakly bond-disordered Potts model, it is generically possible to connect any two spins by a chain of ferromagnetic bonds, even if some are weak, so faraway spins can always indirectly influence each other. But in our system, two faraway $\bm{a}_1$-type domains are typically completely separated by a ``wall'' of $\bm{a}_2$ and $\bm{a}_3$ type domains, so there is no way for them to coordinate their local TR-symmetry ordering and establish long-range order. (Indeed, a calculation shows that in our model, unlike the clock model, domain walls that run parallel to the principal axes have equal energy whether they separate two partially aligned domains (whose order paramers differ by $60^\degree$) or two partially anti-aligned domains (whose order parameters differ by $120^\degree$). Since adjacent domain types have little reason to consistently prefer either partial alignment or partial anti-alignment, two domains of the same type $b$ effectively cannot ``communicate,'' even indirectly, across domains of a different type $b'$.) So even though our disorder model does not \emph{directly} break TR-symmetry as it does the sixfold symmetry, the spin-orbit coupling causes it to indirectly destroy the possibility of spontaneous long-range TR-breaking order via the same mechanism of domain fragmentation -- a phenomenon that does not occur in the RBPM.

The Imry-Ma argument only rules out long-range stripe order, but does not rule out the possibility of inhomogeneous local frozen ordering, i.e.\ a cluster-spin-glass (CSG) phase. A rough physical picture for a state in this phase would have each domain independently freezing into a local value of $p$, which would not fluctuate over time. A transition to such a phase would not be captured by either of our choices of homogeneous order parameter $\langle \psi(x) \rangle$ or $\langle \psi(x)^3 \rangle$. Such a transition would not be visible in our heat capacity measurements either, because the heat capacity per site does not diverge. But while the issue is not completely settled, short-ranged two-dimensional disordered classical magnets do not appear to support finite-temperature CSG phases \cite{Lemke, Parisi}, so in our model the value of $p$ probably does fluctuate over time within each domain. However, recent work has studied the possibility of both CSG and more exotic ``valence-bond-glass'' order in the quantum ground state of \Yb\ \cite{Kimchi} and closely related materials \cite{Ma}. Other recent work has considered a similar bond-disordered spin-orbit-coupled model with six degenerate classical ground states on the 3D pyrochlore lattice, and found a CSG phase over a quite wide range of intermediate disorder strength, surrounded by a long-range-ordered phase at weak (but positive) disorder and a completely unordered paramagnet at strong disorder \cite{Andrade}.

\section{Conclusion \label{Conclusion}}

Our Monte Carlo simulations of the classical version of the model \eqref{Ham} for \Yb\ on quite large systems indicate a single ordering transition from each spin-orbit-coupled phase,  at which the sixfold symmetry is spontaneously broken into stripe order -- although the correlation lengths and correspondingly the finite-size effects are very large -- in agreement with DMRG simulations indicating that the quantum ground state of the clean system is also stripe-ordered \cite{Zhu17}. The ordering is clearly reflected in the spin structure factors and should be detectable by scattering experiments. The single-peaked energy distributions and slow divergences of the heat capacity suggest that the transitions are continuous but not smooth (i.e.\ they are finite-order rather than Kosterlitz-Thouless-like). But we do not know of any non-multicritical 2D CFT that could describe the systems' apparent single critical point, and a field-theory argument seems to indicate that the transitions should be first-order, so we are unable to identify the transitions' universality class.

Adding weak time-reversal-symmetry-preserving bond disorder removes any finite-temperature ordering transition, again in agreement with DMRG results for the quantum ground state (which considered much stronger disorder) \cite{Zhu17}. The Imry-Ma argument shows that infinitesimal bond disorder destroys the possibility of spontaneous long-range stripe order by fragmenting the ground state into (extremely large) domains. We extend the argument to claim that in the spin-orbit-coupled case, the same domain fragmentation mechanism also indirectly prevents the possibility of spontaneously breaking the remaining time-reversal symmetry which survives the addition of disorder. This result supports both the validity of our disorder model for the material \Yb, and Ref.~\onlinecite{Zhu17}'s case that disorder effects rather than spin-liquid physics may explain the observed lack of ordering in the material.

\begin{acknowledgments}
The authors thank John Cardy and Jason Iaconis for helpful discussions and Eric Andrade, Itamar Kimchi, and Sasha Chernyshev for bringing relevant literature to their attention. This work was supported by the DOE Office of Science's Basic Energy Sciences program under Award No.\ DE-FG02-08ER46524. We also acknowledge support from the Center for Scientific Computing at the California NanoSystems Institute and Materials Research Laboratory, an NSF MRSEC (DMR-1720256).
\end{acknowledgments}

\bibliography{Biblio}
\bibliographystyle{apsrev4-1}

\appendix

\section{Classical ground state \label{GS app}}
The Luttinger-Tisza theorem can be used to solve certain classical spin systems of the form
\beq
H = \sum_{i, j} \sum_{\mu, \nu} J^{\mu \nu}_{ij} S_i^\mu S_j^\nu, \label{GeneralH}
\eeq
where $i$ and $j$ range over lattice sites $1, \dots, N$, $\mu$ and $\nu$ over the $\{ x, y, z \}$ directions in spin space, and $J_{ij}^{\mu \nu} \equiv J^{\mu \nu}(i-j)$ is translationally invariant. The idea is to relax the extensive number of normalization constraints $\bm{S}_i \cdot \bm{S}_i = 1$ (no sum on $i$) to the weaker single constraint that the spin norms \emph{average} to $1$, i.e.
\beq
\frac{1}{N} \sum_{i = 1}^N \bm{S}_i \cdot \bm{S}_i = 1, \label{constraint}
\eeq
and minimize the Hamiltonian subject to this looser constraint. If each spin in the resulting formal spin configuration happens to be correctly normalized (which is sometimes but not always the case), then the configuration is the exact ground state of the original problem.

Extremizing the Hamiltonian \eqref{GeneralH} subject to the constraint \eqref{constraint} gives the eigenvalue equation
\beq
\sum_{j, \nu} J_{ij}^{\mu \nu} S_{j}^\nu = \lambda S_{i}^\mu \label{LTeq}
\eeq
(where $\lambda$ is a Lagrange multiplier), which is still nontrivial to solve in general. Typically one makes the Luttinger-Tisza ansatz
\beq
S_{\bm{r}}^\mu = A^\mu \cos \left( \bm{k} \cdot \bm{r} - \delta^\mu \right). \label{LTAnsatz}
\eeq
If the lattice is a Bravais lattice (one site per unit cell), then under this ansatz, \eqref{LTeq} reduces to the nonlinear equation
\beq
\sum_\nu \tilde{J}^{\mu \nu}(\bm{k}) A^\nu \cos \left( \bm{k} \cdot \bm{r}_i - \delta^\nu \right) = \frac{\lambda}{N} A^\mu \cos \left( \bm{k} \cdot \bm{r}_i - \delta^\mu \right), \label{LTAnsatzEq}
\eeq
where
\[
\tilde{J}^{\mu \nu}(\bm{k}) := \frac{1}{N} \sum_j J_{ij}^{\mu\nu} e^{-i \bm{k} \cdot \bm{r}_j}
\]
is the Fourier transform of $J_{ij}^{\mu\nu}$. If the Hamiltonian has a $\text{U}(1)$ spin-space symmetry, then for directions $\mu$ and $\nu$ in the isotropic subspace, $J^{\mu \nu}(k) = J(k) \delta^{\mu \nu}$ and $\lambda = N \tilde{J}^{\mu \nu}(\bm{k})$ and we can always find phase offsets $\delta^\nu$ that normalize all the $\bm{S}_i$. Another tractable special case is when $\delta^\mu \equiv \delta$, in which case \eqref{LTAnsatzEq} reduces to the eigenvalue equation for $J^{\mu \nu}(\bm{k})$. But in this case, the ansatz \eqref{LTAnsatz} is only correctly normalized if the lowest eigenvalue of $\tilde{J}^{\mu \nu}$ is minimized at a commensurate point in the Brillouin zone.

For our system, at the $XXZ$ point on the phase diagram with the SO couplings set to zero, $\tilde{J}(\bm{k})$ is minimized at the $K$ points at the corners of the Brillouin zone, corresponding to $120^\degree$ order, with a continuous degree of freedom in the relative phases between the $\delta^\mu$ corresponding to the $\text{U}(1)$ symmetry. Slightly away from the $XXZ$ point, the wave vector $\bm{k}$ that minimizes the lowest eigenvalue of $\tilde{J}^{\mu \nu}$ becomes incommensurate and the Luttinger-Tisza theorem fails.

But for sufficiently large SO couplings (slightly outside the phase boundaries illustrated in Fig.~\ref{GS phase diagram}), the low symmetry of the strong SO interactions ``locks'' the ground-state wave vector into the commensurate $M$ points at the centers of the Brillouin zone edges, corresponding to stripe order. The lowest eigenvalues of $\tilde{J}(\bm{k}_M)$ cross at the transition between the two SO phases. For the ground state of the in-plane SO phase, the bonds within each stripe (with energy $2 J_{\pm \pm} + 2 J_\pm$) are stronger than the bonds between stripes (with energy $J_{\pm \pm} - 2 J_\pm$). For the ground state of the out-of-plane SO phase, the polar angle $\theta$ begins at a value less than $\pi - (1/2) \arccos(1/3) = 144.7^{\degree}$ at the phase boundary $J_{\pm\pm} \approx -J_{z \pm} / (2 \sqrt{2})$ and decreases monotonically to $\pi/2$ with increasing $J_{\pm\pm}$. The bonds within stripes are stronger for small $J_{z\pm}$, but the bonds between stripes are stronger for large $J_{z\pm}$.

The approximate phase boundaries in Fig.~\ref{GS phase diagram} represent the crossings of the three phases' classical ground-state energies. The transitions are all first-order. Table I of Ref.~\onlinecite{Luo} gives the exact closed-form expressions for the classical ground-state energies, phase boundaries, and (for the out-of-plane SO phase) the canting angle out of the plane.

\section{Order parameters \label{OP app}}

The formal complex vector field $\{ \bm{C}_{\bm{r}} \}$ defined by the inner sum in \eqref{Psi2} depends only on which sublattice the site $\bm{r}$ belongs to. In the in-plane SO phase, is it given by
\begin{align*}
&\bm{C}_A = \frac{3}{2} (1, i, 0), &&\bm{C}_B = \left( \frac{1}{2}, -\frac{3}{2} i, 0 \right), \\
&\bm{C}_C = \left( -1 - \frac{\sqrt{3}}{2} i, -\frac{\sqrt{3}}{2}, 0 \right), &&\bm{C}_D = \left( -1 + \frac{\sqrt{3}}{2} i, \frac{\sqrt{3}}{2}, 0 \right),
\end{align*}
and in the out-of-plane SO phase, by
\begin{align*}
\bm{C}_A &= \frac{3}{2} \sin(\theta) (-i, 1, 0),\\
\bm{C}_B &= \left( \frac{3}{2} i \sin \theta, \frac{1}{2} \sin \theta, 2 \cos \theta \right), \\
\bm{C}_C &=  \left( \frac{\sqrt{3}}{2} \sin \theta, \left( -1 - \frac{\sqrt{3}}{2} i \right) \sin \theta, \left( -1 + i \sqrt{3} \right) \cos \theta \right), \\
\bm{C}_D &= \left( -\frac{\sqrt{3}}{2} \sin \theta, \left( -1 + \frac{\sqrt{3}}{2} i \right) \sin \theta, \left( -1 - \sqrt{3} i \right) \cos \theta \right),
\end{align*}
where $\theta$ is defined in \eqref{theta}.

\end{document}